\documentclass[a4paper,fleqn,usenatbib]{mnras}

\usepackage[T1]{fontenc}
\usepackage{ae,aecompl}
\usepackage{float}

\usepackage{graphicx}	% Including figure files
\usepackage{amsmath}	% Advanced maths commands
\usepackage{amssymb}	% Extra maths symbols
\usepackage{subfigure}
\usepackage{hypcap}
\usepackage{footnote}
\makesavenoteenv{tabular}
\makesavenoteenv{table}
\defcitealias{priscilla20}{Paper~I}
\title[Properties of flares on EV\,Lac]{Properties of flares and CMEs on EV\,Lac: Possible erupting filament}

\author[P. Muheki et al.]{
Priscilla Muheki$^{1,2}$\thanks{E-mail: pmuheki@must.ac.ug},
E.W. Guenther$^{2}$\thanks{E-mail: guenther@tls-tautenburg.de},
T. Mutabazi$^{1}$ and
E. Jurua$^{1}$\thanks{E-mail: ejurua@must.ac.ug}
\\
$^{1}$Mbarara University of Science and Technology, P.O Box 1410, Mbarara, Uganda\\
$^{2}$Th\"uringer Landessternwarte Tautenburg, Sternwarte 5, 07778
Tautenburg, Germany
}

\date{Accepted XXX. Received YYY; in original form ZZZ}

\pubyear{2020}

\begin{document}
\label{firstpage}
\pagerange{\pageref{firstpage}--\pageref{lastpage}}
\maketitle

\begin{abstract}
Flares and CMEs are very powerful events in which energetic radiation and particles
are ejected within a short time. These events thus can strongly affect planets that orbit
these stars. This is particularly relevant for planets of M-stars, because these stars
stay active for a long time during their evolution and yet potentially habitable
planets orbit at short distance. Unfortunately, not much is known about the relation
between flares and CMEs in M-stars as only very few CMEs have so far been observed in
M-stars. In order to learn more about flares and CMEs on M-stars we monitored
the active M-star EV\,Lac spectroscopically at high resolution. 
We find 27 flares  with energies between $1.6\times 10^{31}$  and $1.4\times10^{32}$ erg in $\rm H\alpha$ during  127 hours of  spectroscopic monitoring and 49 flares with energies between $6.3\times10^{31}$ and $1.1\times10^{33}$ erg during the 457 hours of TESS observation. Statistical analysis shows that the ratio of the continuum flux in the TESS-band to the energy emitted in $\rm H\alpha$  is $10.408\pm0.026$. Analysis of the spectra shows an increase in the flux of the \ion{He}{II}\,4686\,\AA\ line  during the impulsive phase of some flares. In three large flares, we detect a continuum source with a temperature between 6\,900 and 23\,000 K. In none of the flares we find a clear CME event indicating that these must be very rare in active M-stars. However, in one relatively weak event, we found an asymmetry in the Balmer lines of $\sim 220\,\rm km\,s^{-1}$ which we interpret as a signature of an erupting filament. 

\end{abstract}


%%%%%%%%%%%%%%%%%%%%%%%%%%%%%%%%%%%%%%%%%%%%%%%%%%

%%%%%%%%%%%%%%%%% BODY OF PAPER %%%%%%%%%%%%%%%%%%

\section{Introduction}
On the Sun, flares and coronal mass ejections (CMEs) are often related and are thought to be caused by the same mechanism. 
% For a long time, studies have been done to fully understand the mechanisms that lead to these phenomena. 
Several studies and observations \citep[e.g.][]{shibata11, aschwanden17} confirm that flares are a result of magnetic reconnection leading to the ejection of magnetic energy.  For very energetic and long lasting flares, chances of being associated with a CME are high \citep{compagnino17}.  Flares and CMEs are well correlated to erupting filaments/prominences \citep{yan11}. As discussed by \citet{gopalswamy03}, about $70\%$ of the erupting prominences are associated with CMEs. In most of these events, the prominence material is seen to lag behind the leading edge of the CME. Erupting prominences can occur both in the active regions where they are more clearly visible \citep{jing04} but also in the quiet regions where the emissions are weak since the magnetic field is weaker \citep{zirin88, forbes00}. It is worth noting that the solar CME rates and properties change over the whole solar cycle and explicitly over time \citep{mittal09}.

Flares and CMEs on other stars especially active M dwarfs are of particular interest. This is because these stars stay active for a long time during their evolution and  potentially habitable planets orbit at short  distances \citep{dressing15}. Notably, these stars show frequent energetic flaring as compared to the Sun \citep{vida17, vida2019, howard19, gunther20}. It has therefore long been thought that they should correspondingly have numerous and powerful CMEs. 
% are presumed to be potential hosts of habitable exoplanets given their vast numbers \citep{dressing15, tuomi19}potentially habitable planets orbit at a shorter distance \citep{dressing13, tuomi19}.have not only been observed on the Sun but also on other stars. Of interest are active .
% % and small radii which enable detection and characterisation of atmospheres of planets at short orbits  
% However, these stars show frequent energetic flaring as compared to the Sun \citep{vida17, vida2019, howard19, gunther20}. It has therefore long been thought that these stars should correspondingly have numerous and powerful CMEs. 
%In such a scenario, flares and CMEs are more relevant .

As we discuss in the following, it is  very important to understand the flare and CME properties on M dwarfs because they play a very important role in the planetary evolution and habitability. Flares are thought to influence prebiotic chemistry or even initiate photosynthesis on planets \citep{buccino06, airapetian16, airapet16,  Mullan2018, rimmer18}. 

On the contrary, several authors \citep[e.g.][]{segura10, venot16, tilley17, vida17,howard19} have shown that flares can  affect the habitability of planets and that the impact of very large events is far more important than many small ones.
% , a number of authors have also studied the negative impact of flares on the habitability of planets \citep[e.g.][]{segura10, venot16, tilley17, vida17,howard19}. 
% The conclusion is that the impact of very large events is far more important than many small ones. 
For example \citet{venot16} examined the influence of the flare on AD\,Leo in 1985 \citep{hawley91} that released $10^{34}$ erg in the wavelength region 120\,-\,800\,nm on the habitability of hypothetical planets around this star. They concluded that it would take about 30\,000 years until the atmosphere would return to its normal state when exposed to a superflare.
In \citet[][hereafter Paper I]{priscilla20}, we found that such energetic flares may occur very frequently ($\sim$ 1 event per hour) on AD\,Leo. This means that the atmosphere of a planet orbiting such a star would be in a state of constant change.
%never be in a normal state but the whole photochemistry of that planet would be changed.
As pointed out by \citet{vida17}, such eruptions can erode planetary atmospheres on the long term, and also directly harm life on the surface. On a brighter note, \citet{abrevaya20} show that there is chance for survival for a certain fraction of the organisms.

Unlike for flares, detections of CMEs on these stars are sparse. Only a handful of events on M dwarfs have been successfully confirmed as CMEs; in which the velocities of the events were larger than the escape velocities of the stars as detected from the enhancements in the blue wings of the Balmer lines \citep{houdebine90, vida16}.
% When studying CMEs we have to keep in mind that blue and red asymmetries are often observed in flares. Although the exact origin is not fully understood, it is likely that they are caused
% by upwards and downward moving material that does not leave the star, because the associated velocities are much smaller than the escape velocities \citep{crespo06, leitzinger10, honda18, vida19, priscilla20}.
% In order to predict the CME occurence rates on M dwarfs, several authors \citep[e.g.][]{leitzinger14, vida16, odert17} have therefore developed models using the stellar flare rates and the solar flare--CME relations. However, there is still a controversy on whether these relations can be extended to M dwarfs given that their activity levels and magnetic field orientations differ from the Sun. 
On the other hand, understanding the dynamics of CMEs on these stars is of paramount importance in the search for possibly habitable exoplanets. These events present a possible threat to habitability since increased CME activity may enhance planetary atmospheric losses through ion pick up mechanisms \citep{lammer08}. Observations of CMEs are therefore important in order to constrain the prediction models \citep{leitzinger14,odert17}.
%hence giving better insights on the CME dynamics on these stars.
% Therefore a better strategy would be to obtain observations of CMEs from these stars and then use the observed rates to predict the occurence rates of these events. 
Since these events are random, observing  active stars for a long time increases the chances of detection. In addition to this, high resolution spectroscopic observations allow for the detection of even the subtle changes in the line profiles of spectra of these objects.

In this paper, we study the properties of flares and search for CMEs on the active M3.5 dwarf, EV\,Lac  which is viewed equator-on ($60^{\circ}$) \citep{morin08} as opposed to AD\,Leo which is viewed nearly pole-on \citepalias{priscilla20}.
% using high resolution spectroscopic observations taken over a long period of time. 
%Both stars have a similar activity rate however are viewed differently on Earth.
EV\,Lac has been a subject of interest in several flare studies on M dwarfs due to its high activity rate.
% In addition to this, it is paramount to establish whether the magnetic field orientation of these stars could act as a constraint in being able to observe CMEs from these stars.
% For this study we selected the active M3.5 dwarf EV\,Lac (=GJ 873) which has two strong spots located at opposite longitudes and with a spot of azimuthal field \citep{morin08}. This star has been a subject of study for a number of flare studies. 

Photometric studies of EV\,Lac by \citet{leto97} revealed that approximately 4.8 flares occur per day on this star. They also suggested that the flaring activity of this star could be  concentrated on a particular longitude.
% \citet{melikian06} did a spectroscopic study of the flares on this star in which they detected a powerful flare that showed a strong blue continuum emission. They observed variations in the equivalent widths of the $\rm H{\alpha}$ and $\rm H{\beta}$ lines which showed a minimum at the flare peak and a maximum 40 minutes later.
\citet{osten10} observed a superflare on EV\,Lac in the Gamma-ray/ X-ray regime whose flux was 7000 times larger than the star's quiescent coronal flux along side variability of the Fe K line which is not found in normal stars. They suggested this as evidence for the existence
of nonthermal hard X-ray emission during flares and used the $\rm K\alpha$ emission to model the flaring loops obtaining loop heights $\rm \sim 0.1\,R_{\star}$. Time-resolved spectroscopy (R\,$\sim$\,10\,000) of a flare on EV\,Lac was done by \citet{honda18}. They found an asymmetry in the blue wing of the $\rm H{\alpha}$ line in each phase of the flare and an absorption feature in the red wing. They suggested that the red absorption component was a result of plasma downflows in the post flare loops. However, the blue asymmetry could not be explicitly explained due to its occurence over the whole flare duration. 
\citet{vida19} also did a statistical study of the occurence rates of CMEs on M dwarfs using archival data of 25 objects including EV\,Lac. They found five slow, weak blue wing enhancements and two stronger events with the strongest having a maximum projected velocity of $513\rm\, km\,s^{-1}$. 

% From these and many other studies, it is perceptible that different aspects about flares emerge and that EV\,Lac is a good target for studies on the flare and CME activity of M dwarfs. However, 
In order to get more insights on the flare and CME dynamics on active M dwarfs, continuous and persistent observations are important. Here we present findings from the high resolution spectroscopic observations of EV\,Lac taken between 2016 and 2019.
%give us more insights on the energetic phenomena on this and other active M dwarfs.\\ 
In addition to the spectroscopic observations, we aim to investigate flares on this star using the high signal to noise data obtained with the Transiting Exoplanet Survey Satellite (TESS) mission.

% With the establishment of the Transiting Exoplanet Survey Satellite (TESS) mission, it is possible to get more insight on flares in M dwarfs since the data has a high signal-to-noise ratio \citep{gunther20}. \\
% In an effort to detect CMEs on these stars, it is important and even more helpful to have a good spectral resolution so that even the subtle changes in the lines can easily be resolved. 
% Using high resolution spectra taken over a long observing period, we were able to detect much smaller wavelength shifts which may give us a better understanding of the CME dynamics on this star.

\section{Observations}
%\subsection{Target}

\subsection{Spectroscopic observations and data reduction}
EV\,Lac was monitored during the flare-search
program of the Th\"uringer Landessternwarte using the 2-m Alfred Jensch telescope of the Th\"uringer Landessternwarte Tautenburg. We use the \'echelle spectrograph with a $2^{\prime\prime }$ slit yielding $\rm R \sim 35\,000$. The spectra cover the wavelength region from 4536 to 7592 \AA . Spectra were obtained in various campaigns from 2016-10-10 until 2019-09-04  with an exposure time of 600\,s. Observations were typically scheduled for one week per month when the object was visible. In 31 nights of obeservations, we obtained 762 spectra in a total monitoring time of 127 hours. The observation log is presented in Table\,\ref{tab1}.

The spectra were bias-subtracted, flat-fielded, corrected for scattered light and extracted using standard Image Reduction Analysis Facility (\textsc{iraf})\footnote{\textsc{iraf} is distributed by the National Optical Astronomy
Observatories, which are operated by the Association of
Universities for Research in Astronomy, Inc., under cooperative agreement with the National Science Foundation.} routines. Wavelength calibration was performed using spectra taken with a ThAr-lamp.

\subsection{Photometric observations}

EV\,Lac was also observed by TESS in Sector 16 (2019-09-11 to 2019-10-16) for 457 hours and  the calibrated, two-minute cadence simple aperture photometry (SAP) data were downloaded from the Mikulski Archive for Space Telescopes (MAST) site\footnote{\url{https://archive.stsci.edu/tess/}}. With its four 10\,cm optical telescopes, TESS simultaneously observes a total field of $24^\circ \times\, 96^\circ$ and each camera has four $\rm 2k\times2k$ CCDs with a pixel scale of 21$^{\prime\prime}$. The telescope's detectors are sensitive in the 600\,-\,1000\,nm wavelength range. This matches part of the wavelength range of our spectroscopic observations e.g. $\rm H{\alpha}$. 
\section{Results and Analysis}
\subsection{Spectroscopic observations}
\subsubsection{Light curves and flare identification}
\begin{figure}

 \includegraphics[width=0.52\textwidth]{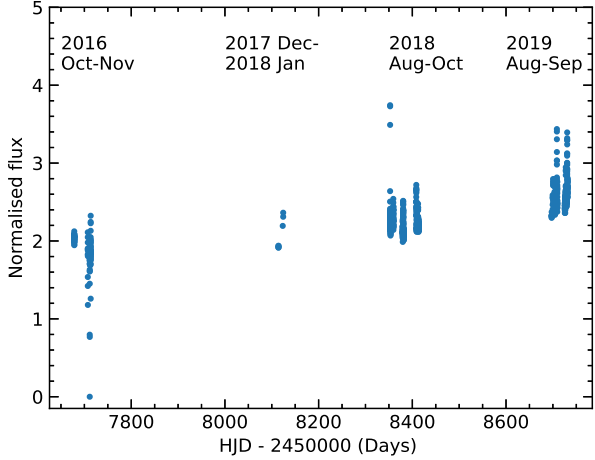}
 
\caption{\label{fig1}Variation of the $\rm H{\alpha}$ flux of EV\,Lac over the entire period of spectroscopic observations.}
\end{figure}

We studied the behaviour of three emission lines; $\rm H{\alpha}$, $\rm H{\beta}$ and \ion{He}{I}\, 5876\,\AA\ (\ion{He}{I}\,$\rm D_3$) from the spectra of EV\,Lac. The flux variation in H$\alpha$ over the entire period of observations is shown in Fig.\,\ref{fig1}. The flux of the line was calculated using the \textsc{sbands} package of \textsc{iraf}. Three passbands were used: one for the line, and two for the continua left and right of the $\rm H\alpha$ line. For the line, a passband width of 4\,\AA\ was used. The $\rm H\alpha$ line flux was then normalised to the mean of the continuum flux.
% In order to construct the lightcurve in Fig.\,\ref{fig1}, we used the measured fluxes in the $\rm H{\alpha}$ line.
From Fig.\,\ref{fig1}, the lightcurve is seen to increase over the observation period. This implies that we observed EV\,Lac during a minimum and towards the maximum of the activity cycle. The lightcurve is characterised by a lower number of flares in the early phase and the number increases from around HJD = 2458400 (August 2018) till the end of our observing campaigns. 

\citet{mavridis86} proposed a 5-year activity cycle for EV\,Lac from the long term fluctuation of its quiescent luminosity. Based on this, we predict that EV\,Lac was at the maximum of its activity cycle in 2019, which explains the higher frequency of flares as compared to the rest of the observing time. 

\begin{figure*}
 \subfigure
{\hspace{-0.2cm}	
\includegraphics[scale=0.49]{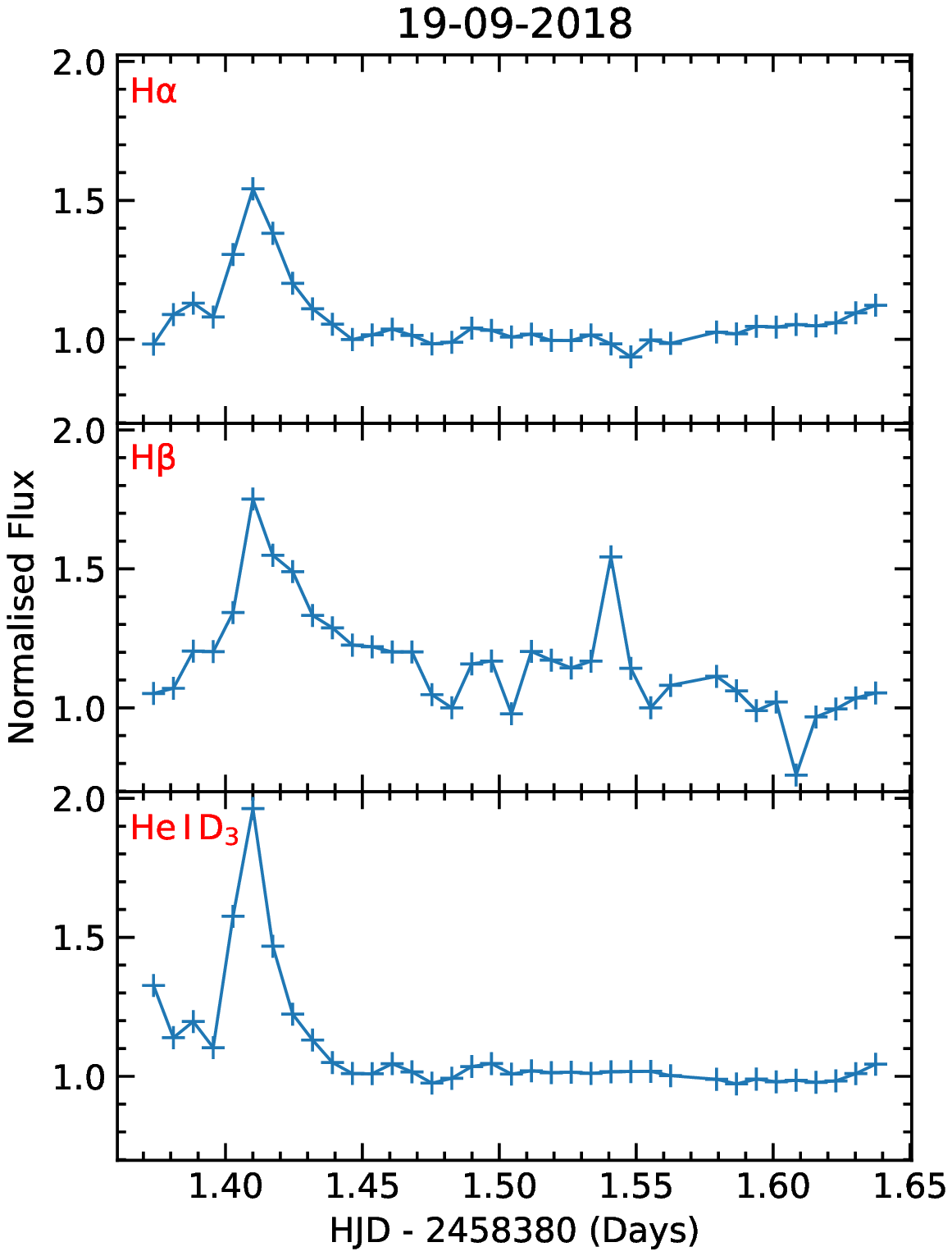}}
\subfigure
{\hspace{-0.2cm}	
\includegraphics[scale=0.49]{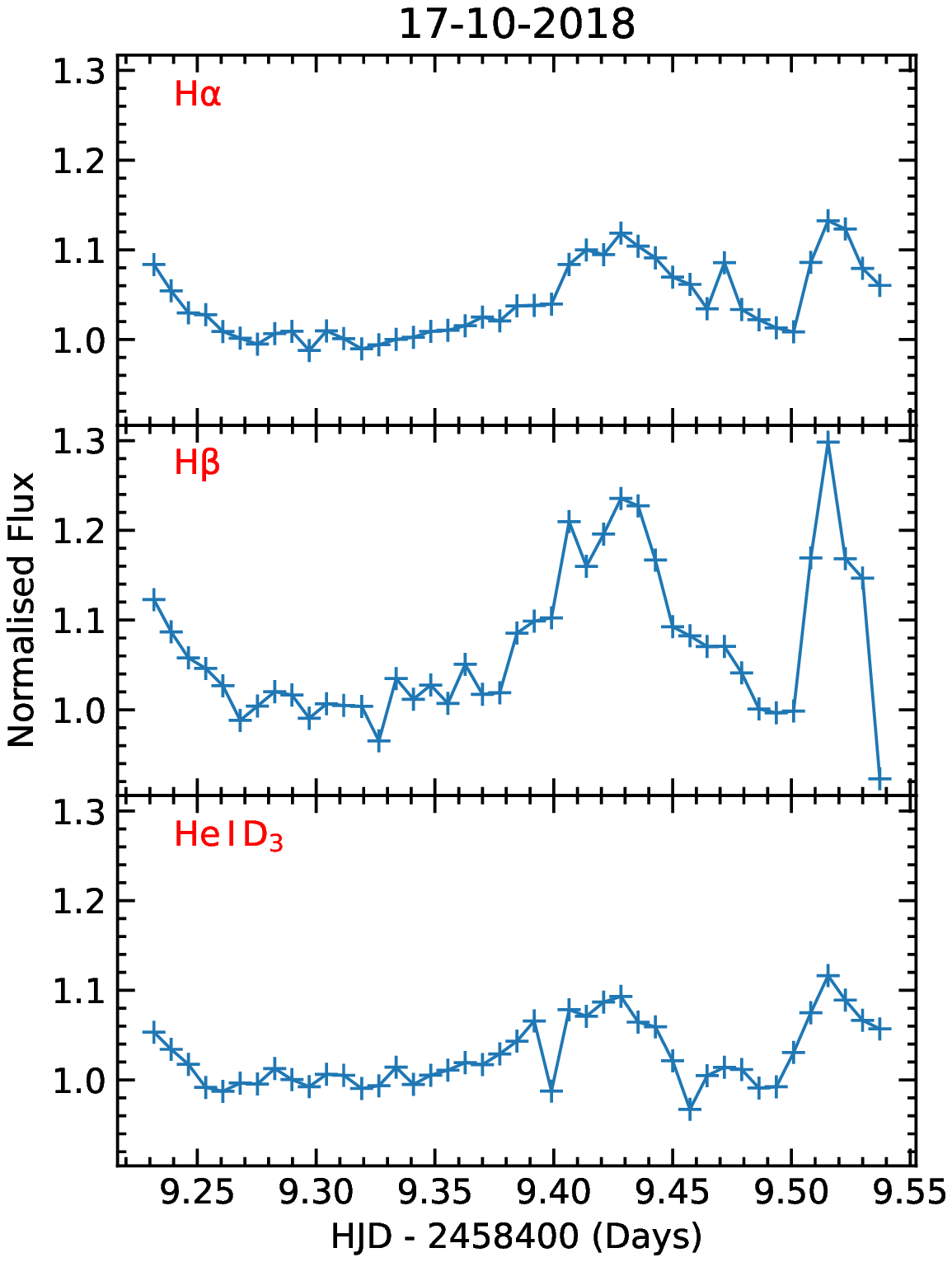}}
% \vskip -0.2cm
\subfigure
{\hspace{-0.2cm}	\includegraphics[scale=0.49]{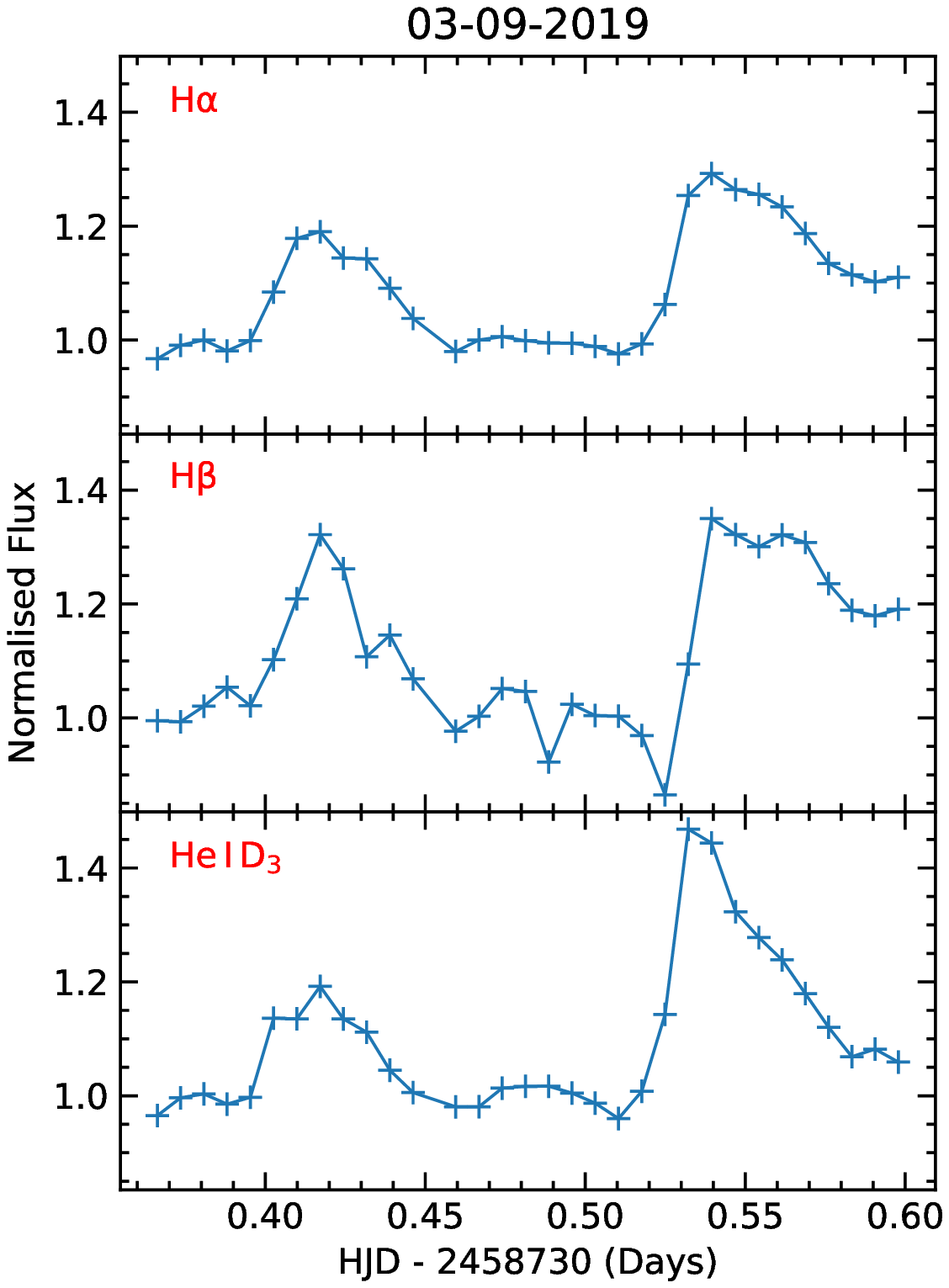}}

\caption{\label{fig2}Temporal evolution of flux in the lines showing lightcurves of some of the flares. The top panels show the evolution of $\rm H{\alpha}$ flux, the middle for $\rm H{\beta}$ and the bottom for \ion{He}{I}\,$\rm D_3$ of that night. }
\end{figure*}

Figure\,\ref{fig2} shows examples of the relative flux variation in $\rm H{\alpha}$, $\rm H{\beta}$ and \ion{He}{I}\,$\rm D_3$ line of some of the flares. As often observed, the flare lightcurves do not follow a similar trend of a usually fast increase and gradual decay but rather being complex with multiple variations during the decay or even a slow impulsive phase often referred to as slow flares as we shall discuss in section \ref{lightcurves}. As indicated in Table \ref{tab2}, in several nights we observed flares only during their impulsive or decay phase. We detected 27 flares in $\rm H{\alpha}$ in the 127 hours of observation which translates to a flare rate of 0.21 flares per hour (5.1 flares per day) with energies $>  1.611\pm0.044\times 10^{31}$ erg. However, it is worth noting that the flare rates depend largely on the activity cycle and therefore may not necessarily explicitly describe the average activity rate of the star.
\subsubsection{White-light flares}
\begin{figure}
\subfigure
   {\includegraphics[scale=0.5]{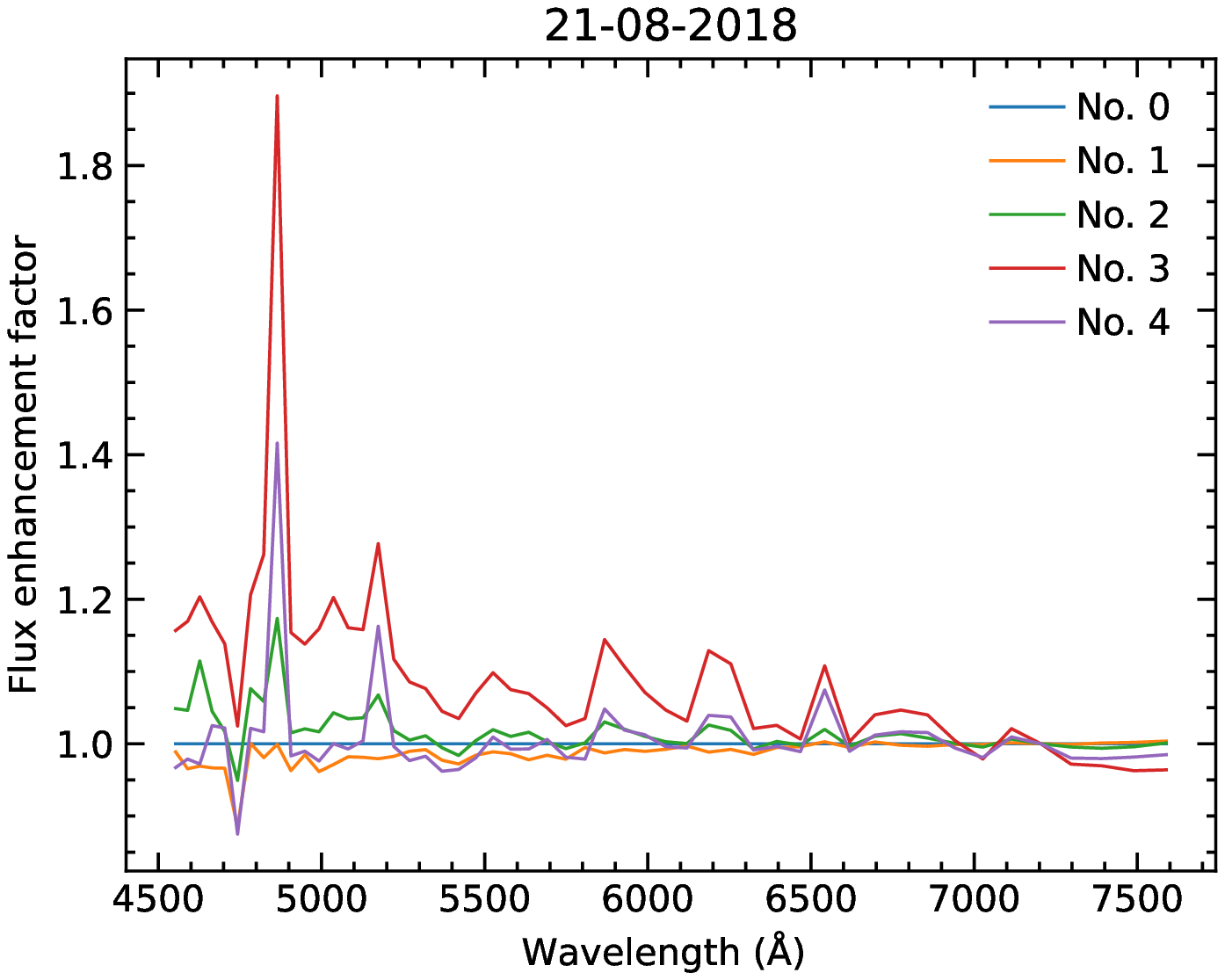}} 
 \vskip -0.29cm
  \subfigure
{\includegraphics[scale=0.5]{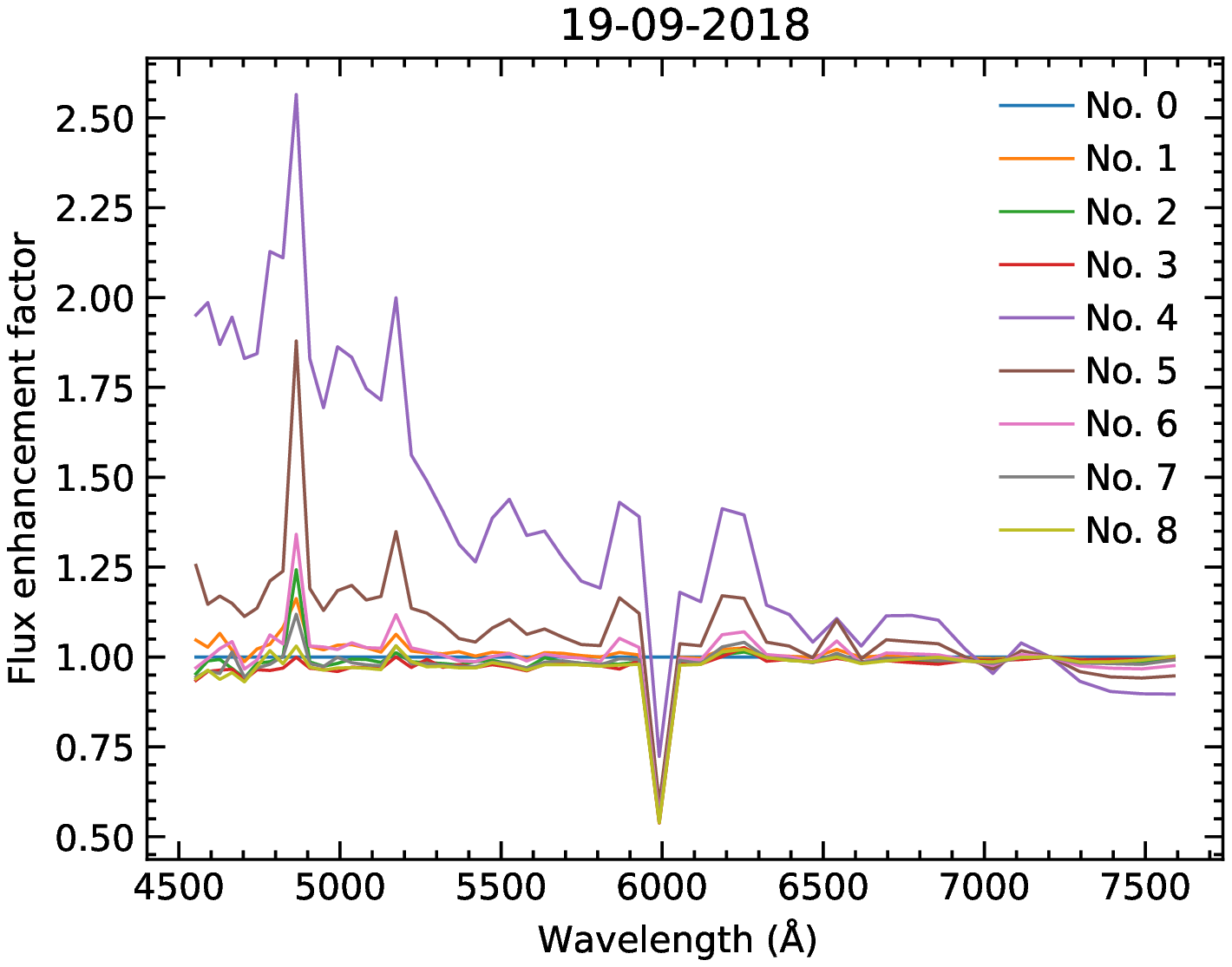}}
\vskip -0.29cm
 \subfigure
 {\includegraphics[scale=0.5]{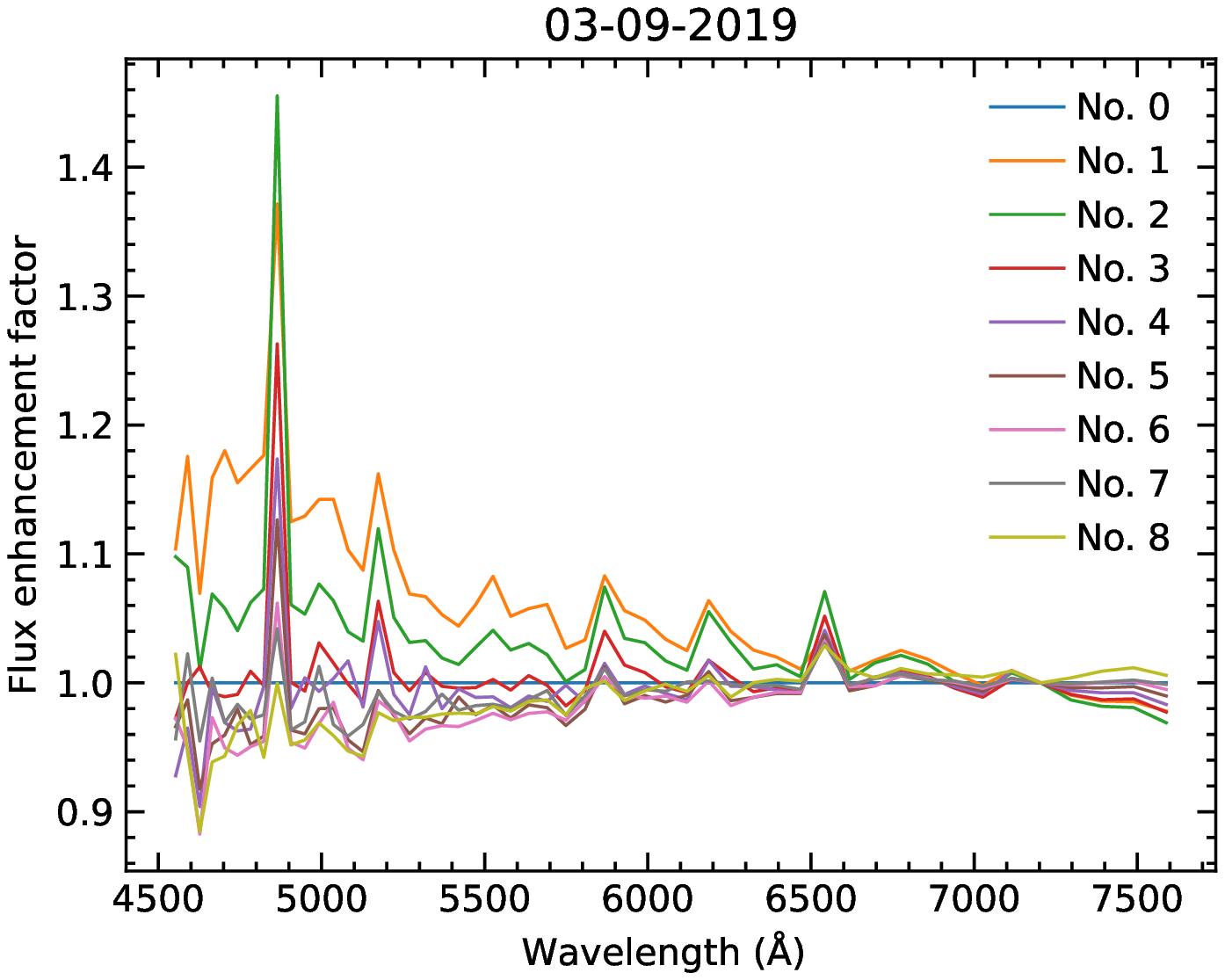}}

\caption{Evolution of the spectra showing the
enhancement in the continuum at the impulsive phase of the flares for F1 (top panel), F2 (middle panel) and F3 (bottom panel).
The numbers in the legend indicate the spectrum numbers where spectrum  No. 0 is the preflare spectrum. }
\label{wlf}

\end{figure}

In this section, we will only discuss the white-light flares observed from spectroscopy. The white-light flares from photometry will be discussed in Section \ref{lightcurves}.
In most of the nights, we observed no significant change in the continuum during the flares and in quiescence. However, on the nights of 2018 August 21, 2018 September 19,  and 2019 September 03 (hereafter, flares F1, F2 and F3 respectively), we observed an enhancement in the continuum at the impulsive phase of the flares and very broadened Balmer lines. These types of flares are known as the Type I white-light flares \citep{machado86}. They are thought to arise due to a deposition of energy in the chromosphere by electron beams producing a white-light continuum by hydrogen recombination \citep{ding96}. The evolution of the white-light flares F1, F2 and F3 is presented in Fig.\,\ref{wlf}.
 
To determine the flux enhancement in the continuum in the spectra, we measured the flux of the star in the centre of each spectral order. These values are a measure of the relative flux of the star in each order.  The continuum emission of flares has typically a temperature of 10\,000 to 20\,000\,K. Since EV\,Lac has a temperature of about 3700\,K \citep{gaia18}, the increase of the continuum brightness due to the
flare is 6 to 10 times larger at 4600\,\AA\ than at 7200\,\AA\,. That
means, the increase is much larger at the short wavelength
than at longer wavelength. By normalising the continuum emission
at a wavelength of 7200\,\AA, we obtain the relative increase
of the continuum flux in the spectrum. We thus measured the
flux of the spectrum at the middle of each spectral order
and normalised these values to 7200 \AA. 
% The Flares on active stars can be described as a blackbody at a temperature of $\approx 10\,000\,\rm K$ and as such change the continuum only very little at longer wavelengths in contrast to shorter wavelengths. We thus decided to normalise the relative fluxes at a wavelength of 7\,200\,\AA. 
One potential source of error of this method is  extinction. However, we observed the change of the continuum brightness in only 3 of the 27 flares and more so during a very brief phase of the flare (see Table \ref{bbt}) and not before or after the event. Since extinction depends on airmass, and we took both pre- and post- flare spectra, we found the effect was very negligible.
%cater for this error in our analysis.
% 6000-7000\,\AAobtained the central wavelength per order of the spectra with blaze with the corresponding mean flux. The normalisation was then done at 7200\,\AA since a blackbody at a temperature of $\approx 10000\,\rm K$ changes the continuum only very little at 6000-7000\,\AA. 
The spectra were then all normalised to the pre-flare spectrum.

In F2, the enhancement was stronger as compared to F1 and F3 and the enhancement is even much more in the blue in agreement with the classification of white-light flares by \citet{machado86}. This kind of white-light flares is seen to occur more frequently on the Sun \citep{ding99} as compared to the Type II events. This blue enhancement can be attributed to the emission by  $\rm H^{-}$ \citep{neidig83, dame85, kleint16}. 
%This implies that both the chromospheric and the photospheric emission contribute to the continuum radiation during flares.

\begin{figure}
\subfigure
{\includegraphics[scale=0.5]{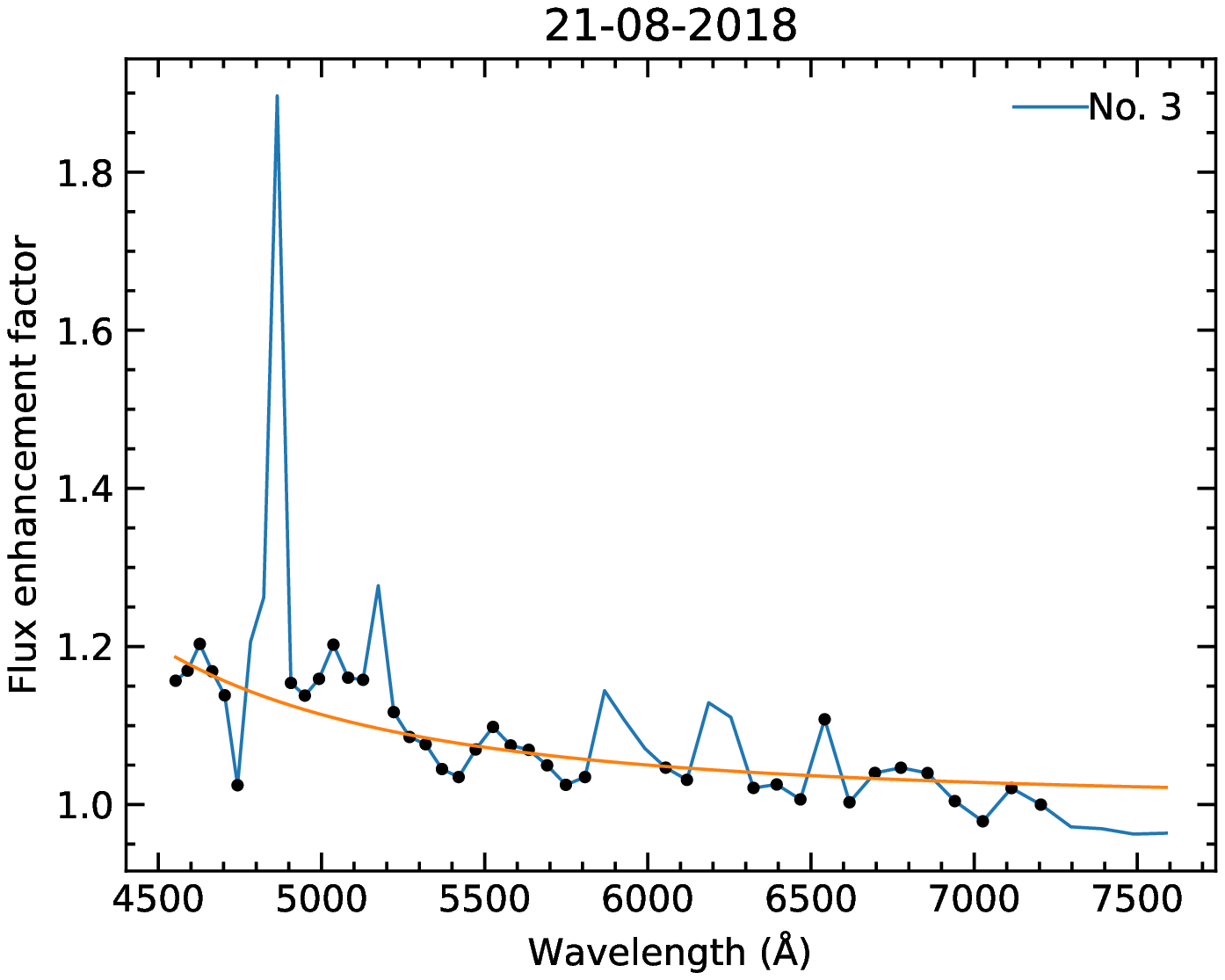}}
\vskip -0.29cm
 \subfigure
{       
\includegraphics[scale=0.5]{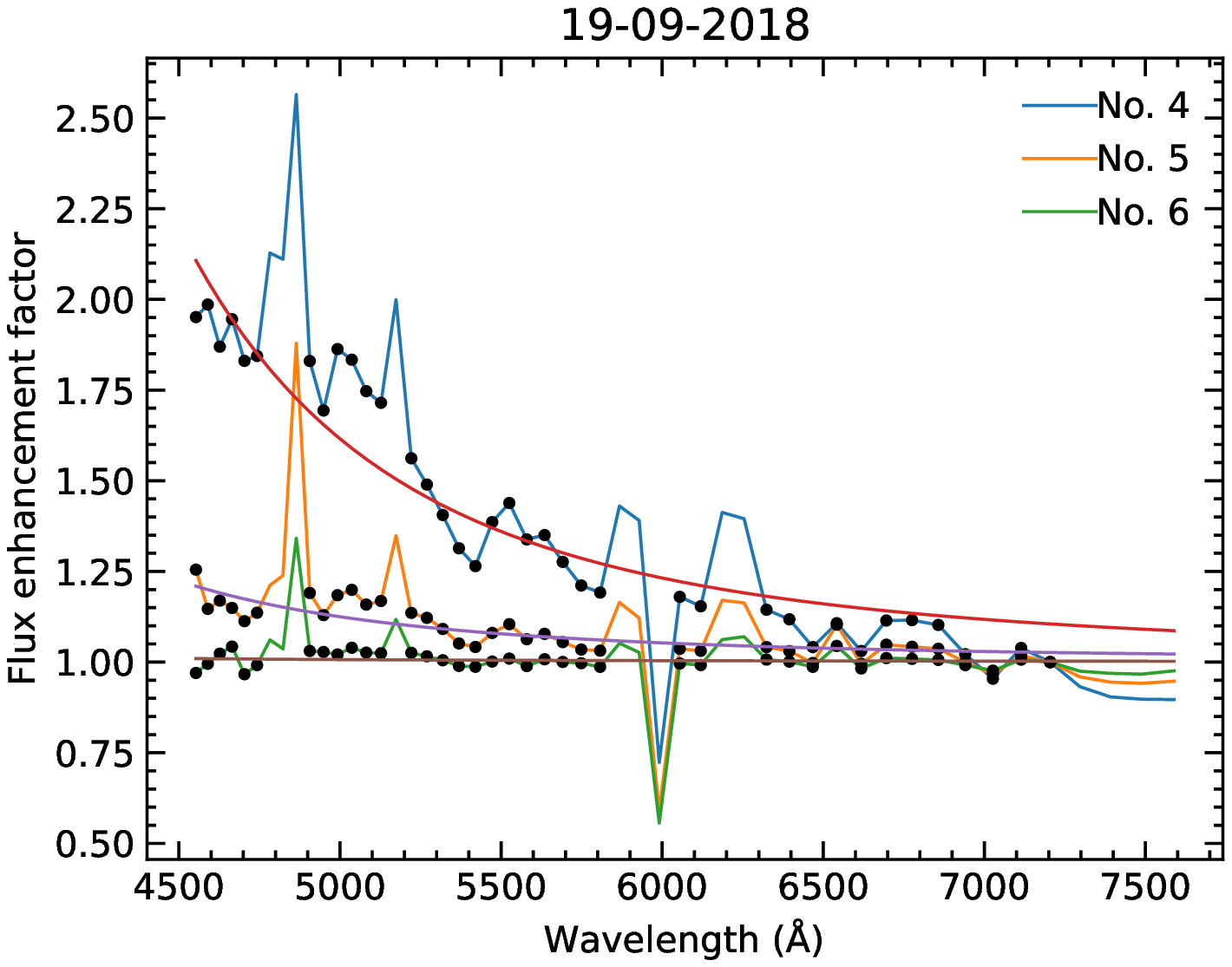}}
\vskip -0.29cm
 \subfigure
 {\includegraphics[scale=0.5]{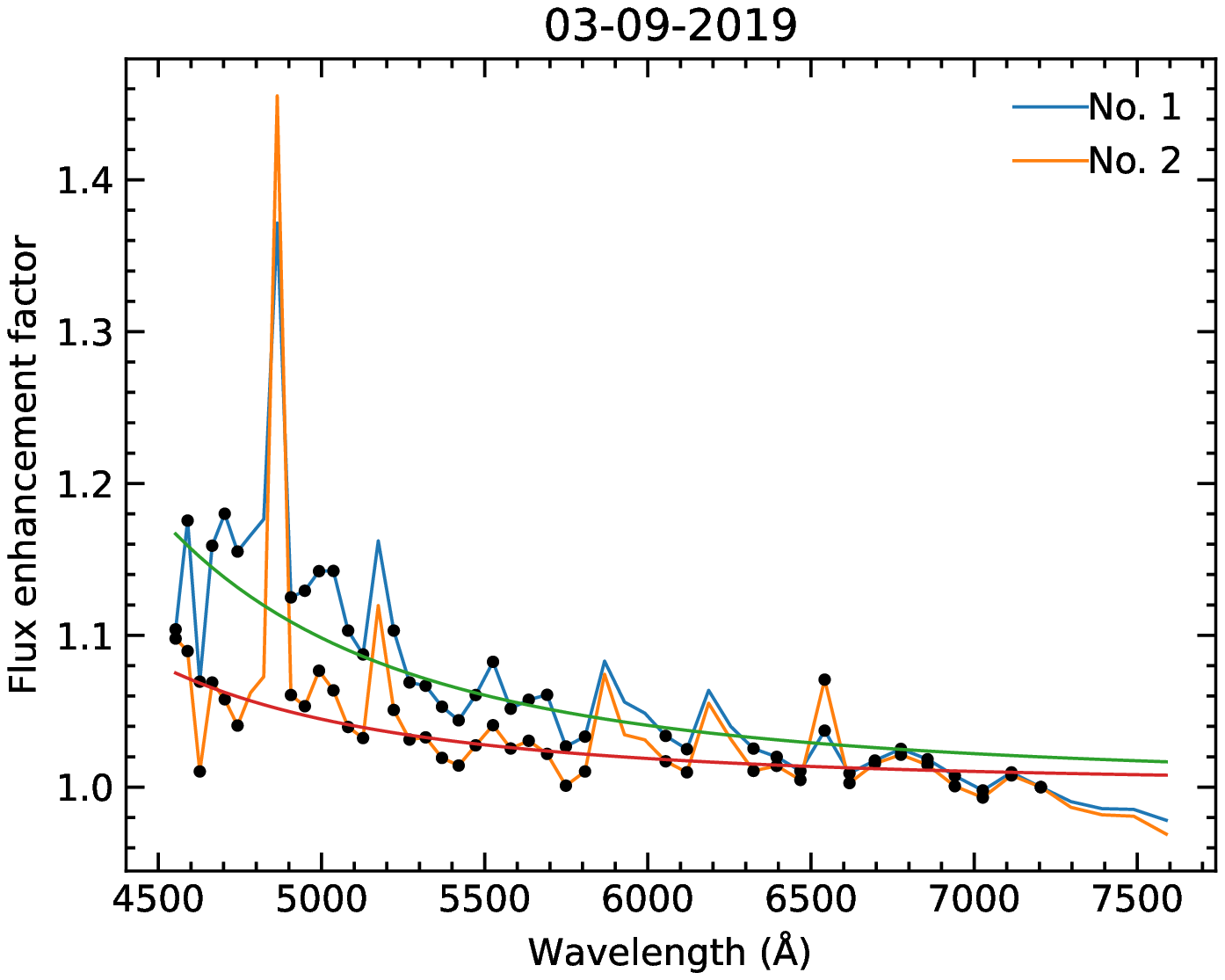}}
%  \vskip -0.35cm
%  \subfigure
% {\includegraphics[scale=0.5]{./Night10lcHe.eps}}
\caption{Blackbody fit to the one spectrum showing continuum enhancement in F1 (top panel), three spectra in F2 (middle panel) and two spectra in F3 (bottom panel). }
\label{bb}

\end{figure}

Because the continuum flux of a flare can be approximated as a blackbody, we used the relative fluxes to measure the temperature of this blackbody. We did this by fitting a blackbody curve to the relative increase of the continuum. Again, because the increase is much larger at shorter wavelengths than at longer wavelengths, also the fit is dominated by the measurements at the shorter wavelength. For this reason, the blackbody curve is always larger than unity at 7200\,\AA\ as shown in Fig.\,\ref{bb}.
We note that there is a peak at about 4800\,\AA\,. This is because there is an emission line close to the centre of that order. However, we do not consider this point in our fit.
% fluxes of one spectrum in the impulsive phase of F1, three spectra at the peak of F2 and two spectra in F3 in which the continuum enhancement was observed (see Fig.\,\ref{bb}).
The values of the temperature of the flare and area fraction obtained for the flares is summarised in Table \ref{bbt}.
\begin{table}
\caption{Summary of the values of temperature and area fraction of the white-light flares.\label{bbt} }
\begin{tabular}{ccccc}
 
   \hline
   \hline
  Flare&Spectrum&${t_o}/{t}$ (min)$^a$&Temp. (K) & Area \\
  &&&&fraction ($\%$)\\
   \hline
   2018-08-21&Impulsive&${10}/{260}$&11\,600&0.057\\
   2018-09-19&Impulsive&${30}/{360}$&23\,000&0.070\\
   &Peak&&13\,100&0.046\\
   &Decay&&6\,900&0.020\\
   2019-09-03&Impulsive&${20}/{320}$&14\,100&0.030\\
   &Peak&&13\,300&0.016\\
  
   \hline
   
  \end{tabular}
  ~\\
   \footnotesize{$^a$ is the duration of the white-light enhancement relative to the duration of the observation of that night.}
  %\caption{Table.1. Summary of the observation log}
  \end{table}
  
% For F2, we obtained a temperature of $\rm \approx 22984 \,K$ and an area fraction of $\approx 0.070\%$, $\rm \approx 13070\, K$ and $0.05\%$ and $\rm \approx 6907\,K$ and $0.020\%$. For F3, the temperature obtained from the fit is $\rm \approx 14178 \,K$ with an area fraction of $\approx 0.03\% $ and $\rm \approx 13317\,K$ covering an area of $\approx 0.02\%$.

% The high temperatures obtained in F2 and F3 could possibly indicate that we observe the reconnection site.
%re indicative of a possibility  that these flares occured at sites where there was magnetic reconnection. 
\subsubsection{Flare energies}
\label{flarene}
The energy released during a flare was estimated by integrating the luminosity over the duration of the flare. In order to determine the absolute line fluxes, we used the flux of each line and its nearby continuum. A flux calibrated spectrum of EV\,Lac was obtained by scaling the flux calibrated spectrum of AD\,Leo from \citet{cincunegui04} to the flux level of EV\,Lac using the \textit{J}, \textit{H} and \textit{K} brightness measurements given in the SIMBAD database. The flux calibration was done in the same way as in \citetalias{priscilla20}.  %continuum flux near the emission lines  was then determined by direct scaling of this flux calibrated spectrum. 
We assumed a constant continuum for the flares showing no white light enhancement. However, for the white-light flares,  the increase in the continuum was considered.
% this is a lower limit when the continuum increases as is the case for the white-light flares. 
The luminosity was calculated using the flux of the line and the distance of $5.050\pm0.001$\,pc to EV\,Lac obtained from \textit{Gaia} \citep{gaia18}. The energies released in each observed flare are given in Table \ref{tab2} for $\rm H{\alpha}$, $\rm H{\beta}$ and \ion{He}{I}\,$\rm D_3$. In Table \ref{tab2_}, we give the energy of the white-light flares taking into account the continuum increase.

\begin{table*}
\caption{ Summary of the energy of the flares observed in  $\rm H \alpha$, $\rm H \beta$ and \ion{He}{I}\,$\rm D_3$ in columns 2-4. In columns 5 and 6 are the number of spectra with asymmetries in the blue and the red wings respectively. Columns 7 and 9 represent the maximum line-of-sight velocities observed in that flare in the blue and red respectively. In columns 8 and 10, the average of the line-of-sight velocities of the asymmetries in that flare are also presented.}\label{tab2}
  \begin{tabular}{lcccccccccccccc}
\hline
\hline
Date&\multicolumn{3}{c}{Flare Energy ($10^{31}$ erg)}&&\multicolumn{2}{c}{No. of spectra}&&\multicolumn{2}{c}{Vel.-Blue ($\rm km\,s^{-1}$)}&&\multicolumn{2}{c}{Vel.-Red ($\rm km\,s^{-1}$)}\\
&H$_\alpha$&H$_\beta$&\ion{He}{I}\,$\rm D_3$&&Blue&Red&&max.&av.&&max.&av.\\[0.5ex]
\hline
2016-11-19&3.51$\pm$0.05&$1.065\pm0.032$&$0.145\pm0.003$&&&&&&&&&\\
2018-08-21&6.62$\pm$0.07&$1.626\pm0.047$&$0.156\pm0.002$&&4&4&&179&150&&111&108\\
2018-08-21~$^\dagger $(F1)&3.85$\pm$0.30&1.70$\pm$0.13&$0.112\pm0.01$&&4&4&&369&283&&450&276\\
2018-08-22~$^\dagger$&7.20$\pm$0.07&$0.651\pm0.009$&$0.091\pm0.001$&&&&&&&&&\\
2018-08-22~$^\dagger$&$6.082\pm0.048$&$0.544\pm0.008$&$0.141\pm0.002$&&&&&&&&&\\
2018-09-17&$5.283\pm0.049$&$1.346\pm0.025$&$0.125\pm0.001$&&&&&&&&&\\
2018-09-17&$4.622\pm0.051$&$0.618\pm0.001$&$0.079\pm0.001$&&&&&&&&&\\
2018-09-18 &9.50$\pm$0.09&$1.423\pm0.024$&-&&&&&&&&&\\
2018-09-19 (F2)&11.94$\pm$0.51&$2.093\pm0.068$&$0.268\pm0.021$&&8&8&&410&234&&385&218\\
2018-10-17~$^\dagger$&4.21$\pm$0.05&0.78$\pm$0.01&-&&&&&&&&&\\
2018-10-17&12.23$\pm$0.10&$2.170\pm0.037$&$0.267\pm0.003$&&&&&&&&&\\
2018-10-17~$^\dagger$&$3.718\pm0.046$&$0.816\pm0.040$&$0.104\pm0.001$&&&&&&&&&\\
2018-10-20~$^\dagger$&3.17$\pm$0.05&$0.671\pm0.017$&$0.074\pm0.001$&&&&&&&&&\\
2018-10-21&10.45$\pm$0.03&1.87$\pm$0.01&$0.259\pm0.001$&&&&&&&&&\\
2019-08-07~$^\dagger$&5.33$\pm$0.05&$1.218\pm0.018$&$0.102\pm0.001$&&&&&&&&&\\
2019-08-07~$^\dagger$&$6.296\pm0.078$&$1.442\pm0.026$&$0.140\pm0.001$&&&&&&&&&\\
2019-08-08~$^\dagger$&$7.255\pm0.027$&$1.486\pm0.007$&$0.159\pm0.0004$&&&&&&&&&\\
2019-08-13~$^\dagger$&1.61$\pm$0.04&0.37$\pm$0.02&$0.039\pm0.002$&&2&2&&151&148&&160&136\\
2019-08-30&13.71$\pm$0.05&$2.702\pm0.016$&$0.122\pm0.001$&&&&&&&&&\\
2019-08-30~$^\dagger$&$5.758\pm0.038$&$1.409\pm0.022$&$0.290\pm0.001$&&&&&&&&&\\
2019-08-31~$^\dagger$&10.05$\pm$0.14&$1.943\pm0.045$&$0.241\pm0.003$&&&&&&&&&\\
2019-09-02&$3.777\pm0.045$&-&-&&&&&&&&&\\
2019-09-02&6.44$\pm$0.05&-&-&&4&4&&288&194&&88&85\\
2019-09-03&9.03$\pm$0.21&$1.710\pm0.058$&$0.189\pm0.005$&&3&3&&181&151&&109&93\\
2019-09-03~$^\dagger $(F3)&10.84$\pm$0.22&$2.095\pm0.044$&$0.26\pm0.01$&&10&10&&320&187&&280&168\\
\hline
%\hline
\end{tabular}
~\\
$\dagger$ indicates flares in which only the decay or impulsive phase was observed.
\end{table*}

\begin{figure}
 
 \includegraphics[width=.52\textwidth]{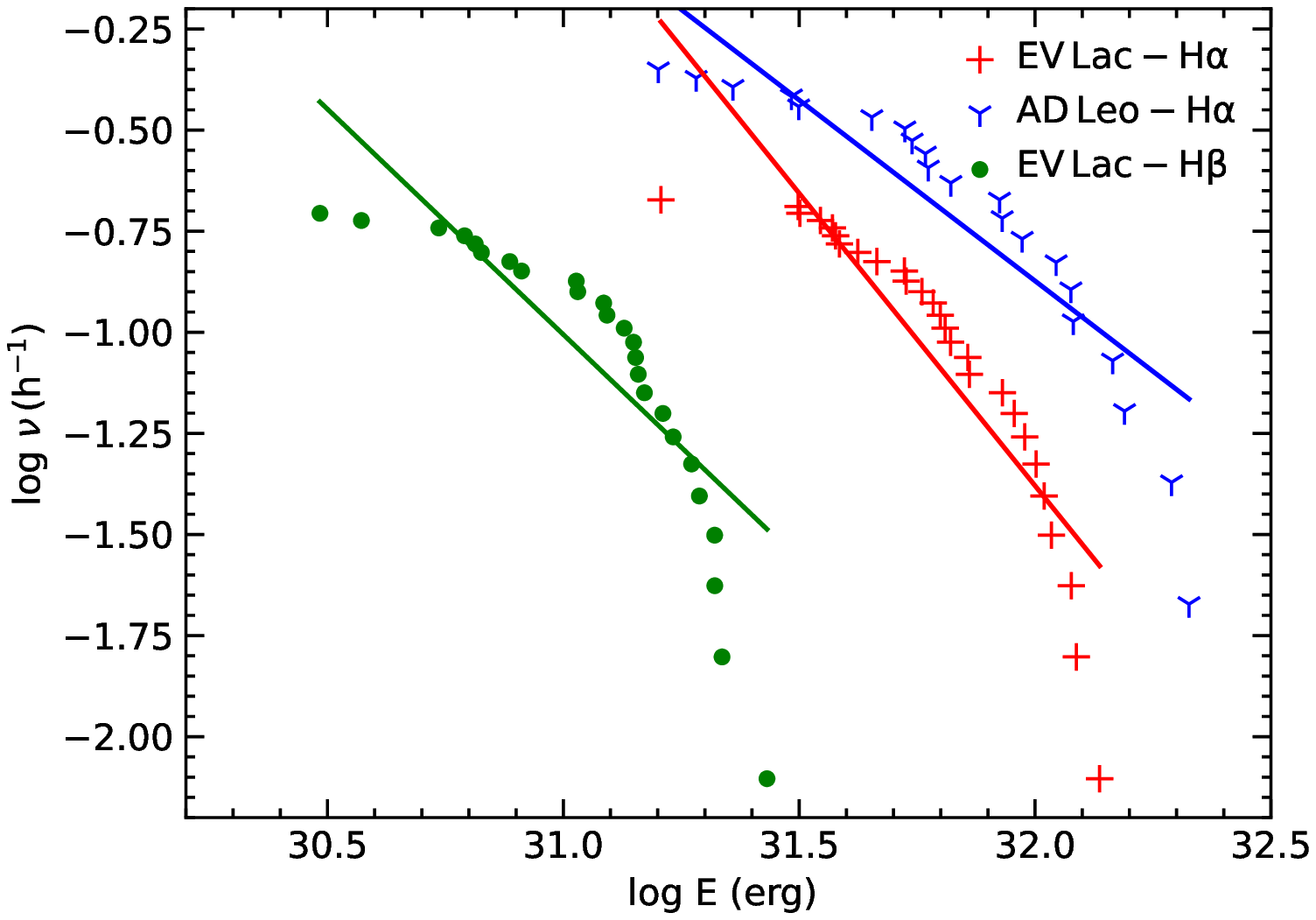}
 \caption{\label{fig3}Cumulative flare frequency distribution of flares on EV Lac in H$_\alpha$ (red plus symbols) and H$_\beta$ (green filled circles). For comparison, AD Leo \citepalias{priscilla20} is also shown in blue tri symbols. The solid lines represent our best linear least-squares fit to each of the Flare Frequency Distributions.}
\end{figure}

%\bigskip
\begin{table}
 \caption{Flare energies of the white-light flares taking the continuum increase into account.}\label{tab2_}
 \begin{tabular} {p{0.5cm} p{0.5cm} p{2.0cm} p{2.0cm} p{2.cm} p{1.cm}cccc}

\hline
\hline
Flare &&\multicolumn{3}{c}{Flare Energy ($10^{31}$ erg) }\\
&& $\rm H{\alpha}$ & $\rm H{\beta}$ & \ion{He}{I}\,$\rm D_3$\\[0.5ex]
\hline
F1&&3.86$\pm$0.30&1.11$\pm$0.13&0.11$\pm$0.01\\
%$\rm l.s.f^b$&&$-1.56\pm0.13$&$-1.10\pm0.08$&--\\
%$\rm m.l.e^c$&&$-1.45\pm0.26$&$-1.12\pm0.35$&$-1.01\pm0.44$\\
F2&&12.23$\pm$0.66&2.83$\pm$0.10&$0.275\pm0.024$\\
F3&&10.85$\pm$0.22&$2.278\pm0.054$&$0.261\pm0.009$\\

\hline
\end{tabular}
\end{table}
We calculated the cumulative flare frequency distribution by fitting a cumulative power law to the flares. Since we measured the continuum increase only in three events, the cumulative flare frequency distribution was calculated using the line fluxes without taking the continuum increase into account so that all flares are treated the same way. The cumulative flare frequency distribution follows a power law of the form \citep{lacy76}:
\begin{equation}
 \log \nu = \beta\log E + b,
 \label{ffd}
\end{equation}
 where $\nu$ is the number of flares with an energy equal to or greater than $E$  occuring per hour, $\beta$ is the frequency of occurence of flares with various energies and $b$ the y-intercept which sets the flaring rate.
 The flare frequency distribution can also be expressed as:
 \begin{equation}
  dN \propto E^{-\alpha} dE,
  \label{ffd1}
 \end{equation}
 where $dN$ is the number of flares with energies in the range  $[E,E+dE]$.
 The slope $\beta$ in Eq.\,\ref{ffd} and the index $\alpha$ in Eq.\,\ref{ffd1} are related as $\beta=1-\alpha$.
 By ignoring flares with energy below $5.28\times 10^{31}$ erg and $1.35\times 10^{31}$ erg in $\rm H{\alpha}$ and $\rm H{\beta}$ respectively in our fit, we obtained $\beta = -1.81\pm0.21$ and $\beta = -2.58\pm0.51$ for flares in $\rm H{\alpha}$ and $\rm H{\beta}$ respectively. 
In Fig.\,\ref{fig3}, we  present the cumulative frequency distribution of EV\,Lac. We have included the cumulative distribution frequency for AD\,Leo ($\beta=-1.56\pm0.13$) obtained from \citetalias{priscilla20} for comparison. Comparing the distributions of the two stars, we find that AD\,Leo  is on average more active than EV\,Lac by a factor of $\sim 3$. 
A study of the flare rates in cool stars by \citet{gizis17} indicates that using a linear fit to describe the cumulative flare frequency distribution may yield errors since the distribution is usually tailed. They suggested the use of the maximum likelihood estimation to estimate the parameter $\alpha$. We therefore also applied the maximum likelihood estimation to determine the $\beta$ values since the two indices are related as discussed above. The values obtained are summarised in Table \ref{beta}. The $\rm H\alpha$ energies of the flares are on average a factor $5.27\pm0.40$ higher than in $\rm H\beta$ and a factor $50.92\pm8.23$ higher than  in \ion{He}{I}\,$\rm D_3$. These factors were derived from the energies of the flares whose entire evolution was observed.
 \begin{table}
\caption{Summary of the $\beta$ values obtained by least-squares fitting (l.s.f)  and maximum likelihood estimation (m.l.e). Columns 2 and 3 correspond to the flares from the Balmer lines and column 4 is the white-light flares from TESS.}
  \begin{tabular} {p{0.5cm} p{0.5cm} p{2.0cm} p{2.0cm} p{2.cm} p{1.cm}cccc}
%\hline
\hline
Method &&\multicolumn{3}{c}{$\beta$ value}\\
&& $\rm H{\alpha}$ & $\rm H{\beta}$ & TESS\\[0.5ex]
\hline
$\rm l.s.f$&&$-1.81\pm0.21$&$-2.58\pm0.51$&$-1.78\pm0.17$\\
%$\rm l.s.f^b$&&$-1.56\pm0.13$&$-1.10\pm0.08$&--\\
%$\rm m.l.e^c$&&$-1.45\pm0.26$&$-1.12\pm0.35$&$-1.01\pm0.44$\\
$\rm m.l.e$&&$-1.96\pm0.55$&$-2.71\pm0.94$&$-2.63\pm0.82$\\

\hline
\end{tabular}
~\\ \footnotesize{The value of \textit{b} obtained from Eq.\ref{ffd} is $56.81\pm5.33$, $79.22\pm5.97$ and $56.48\pm4.86$ for $\rm H\alpha$, $\rm H\beta$ and TESS respectively.}
 \label{beta}
 \end{table}

\subsubsection{An erupting quiescent filament and line asymmetries}
\label{ep}
We searched for line asymmetries to detect CMEs. This was done by subtracting an average spectrum in quiescence of each night from all the night's flare spectra. The maximum line-of-sight velocity was then estimated at the point where the residual profile merges with the continuum.  
A clear signal of a CME would be a blue shifted component in the emission line with a line-of-sight velocity greater than the escape velocity of the star. For EV\,Lac, the escape velocity is $\approx 620\, \rm km\,s^{-1}$. 
% We observed asymmetries in the blue and red especially at the flare peak. However, all the detected asymmetries present with velocities lower than  the escape velocity of the star as shown in Table \ref{tab2}. 
% The spectra in the night of 2 September, 2019 (see Fig.\ref{nyt12}) showed a blue asymmetry corresponding to a line-of-sight velocity $\approx 220\, \rm km\,s^{-1}$ during a weak and slow flare which was detected only in $\rm H{\alpha}$. According to \citet{zirin88}, there are prominences which erupt slowly and only a small brightening occurs. These occur in quiet regions where the density of the plasma is lower. These eruptive prominences are responsible for CMEs on the Sun since the strength of the magnetic fields is much lower as compared to the active regions. 

% This could be from an unresolved flare of lower energy \citep{crespo06}. The other possibility could be that it was an effect of a stronger flare which might have taken place earlier \citep{fuhrmeister18}. It is also possible that due to geometrical effects, we observe the material being ejected from an erupting filament loop observed at the horizon but not the flare if it happens on the rear side of the star.

During the flare evolution, the lines are also broadened especially at the impulsive phase and early decay phase of the flare. This could likely be due to Stark broadening effect as a result of the pressure from the electrons \citep{svestka72}. 

 \begin{figure}
  
  \includegraphics[width=.48\textwidth]{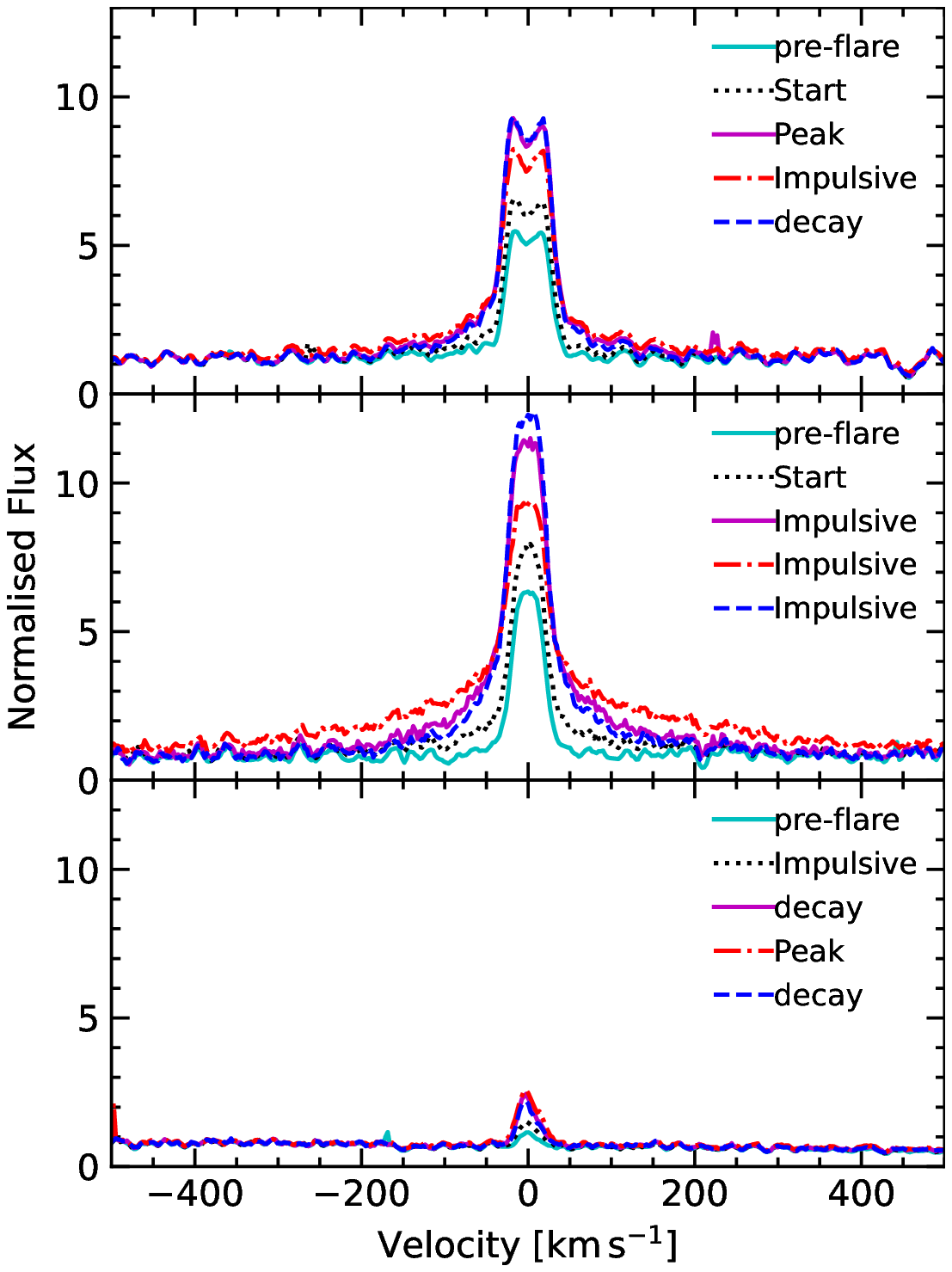}
  \caption{Evolution of the flare F1 showing the asymmetries in the blue and red wings at flare peak in $\rm H{\alpha}$ (top panel), $\rm H{\beta}$ (middle panel) and \ion{He}{I}\,$\rm D_3$ (bottom panel).\label{asy1} }
%  \caption{\label{nyt12} Evolution of the flare F2 showing the asymmetries in the blue and red wings at flare peak.}
 \end{figure}

Figure\,\ref{asy1} shows part of the flare evolution of the white-light flare F1 in $\rm H\alpha$, $\rm H\beta$ and \ion{He}{I}\,$\rm D_3$. This flare shows asymmetries (both in the blue and the red). We observed only the impulsive phase of this flare and it is seen to peak at different times in the different lines. At the impulsive phase in all the lines, a red asymmetry is observed and at the flare peak the lines are symmetrically broadened. The maximum line-of-sight velocities obtained for the different flares  can be found in Table\,\ref{tab2}. The velocities in the blue component obtained in all flares are all far below the escape velocity of the star (cf Table\,\ref{tab2}).
% By fitting a gaussian to the residual profile we obtain a velocity of $\rm \approx 350 km\,s^{-1}$. 

As has been discussed by previous studies \citep[e.g.][]{crespo06, leitzinger14, fuhrmeister18, vida19}, such low velocity asymmetries have been observed on the Sun and other stars. They are thought to originate from flare plasma motions such as chromospheric evaporations and condesations. Observations of asymmetries on the Sun indicate that mass motions can have velocities in the order of a few tens of $\rm km\,s^{-1}$ \citep{tei18} and the most explosive extend to a few hundreds of $\rm km\,s^{-1}$ \citep[e.g.][]{canfield90}. Another possible explanation for the asymmetries is the detection of erupting filaments as have been observed on the Sun. 

CMEs often show a three part structure: the bright front that corresponds to the ejected material at the highest speed, the dark cavity carrying intense magnetic fields and the filament which travels at a lower speed \citep{illing86}. According to \citet{zhang01}, CME eruptions on the Sun follow phases of initiation, acceleration which coincides with the impulsive phase of the associated flare and propagation where a constant speed is observed. The impulsive acceleration of the CME at the impulsive phase can be attributed to the erupting material in the filament \citep{kundu04}. However, a filament is much slower than the CME.
The velocities of such eruptions have been found to lie in the range of few tens to hundreds of $\rm km\,s^{-1}$ e.g. \citet{gopalswamy03} found events with an average speed of $64.5~\rm km\,s^{-1}$  up to $\approx 380\,\rm km\,s^{-1}$ and even higher $ \approx 650\,\rm km\,s^{-1}$ \citep{kundu04}. This range fits well with the velocities obtained in our observations. It is probable that we observe erupting filaments that occur during the early phase of the eruption. 
%This would imply that we do not observe the bright front which corresponds to the fast travelling ejected material. 

\begin{figure}
 
 \includegraphics[width=.52\textwidth]{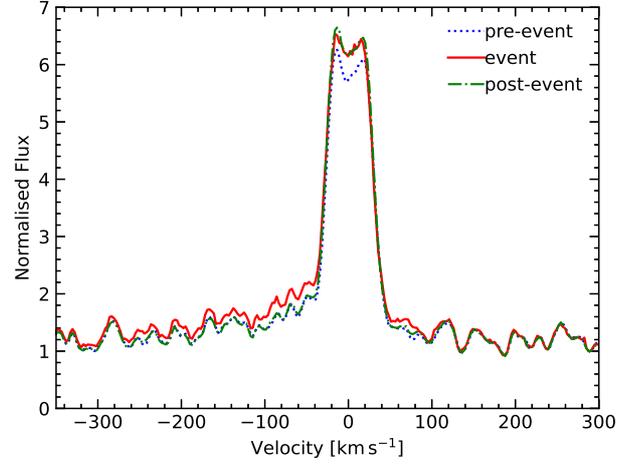}
 \caption{\label{nyt12}One of the spectra of the night in which no typical flare was observed but blue asymmetry in the continuum was observed in $\rm H{\alpha}$.}
\end{figure}

The spectra in the night of 2019 September 2  (see Fig.\,\ref{nyt12}) shows a blue asymmetry corresponding to a line-of-sight velocity $\approx 220\, \rm km\,s^{-1}$ during a weak and slow flare which was detected only in $\rm H{\alpha}$. Such weak flares have an early thermal phase, when a filament is erupting, that may involve some weak $\rm H{\alpha}$ emission \citep{bai89}. 
% It appears that a flare of this class does not exhibit a main phase (either non thermal or thermal), but it does exhibit both the late thermal phase and the late nonthermal phase that we attribute to the slow reconnection of a newly formed  current sheet and to shock waves generated by the eruption. 
% According to \citet{zirin88}, there are prominences which erupt slowly and only a small brightening occurs.
These occur in quiet regions where the density of the plasma and the strength of magnetic fields is lower than in the active regions. This kind of erupting filaments has been found to contribute to a reasonable fraction of CMEs on the Sun \citep{jing04,gosain16}. This implies that CMEs may not always be associated with only very large flares.  We therefore interpret this asymmetry as a signature of an erupting quiescent filament. However, it should still be noted that we did not detect the CME that could have been associated with it.
\subsubsection{Detection of the \ion{He}{II}\,4686\,\AA\ line}
In two of the spectra in F2 (indicated as spectrum No.\,4 and No.\,5 in the middle panel in Fig.\,\ref{wlf}) and in one spectrum (No.\,2 in the bottom panel of Fig.\,\ref{wlf}) in F3 we detected the \ion{He}{II}\,4686\,\AA\ line which has only been previously detected in hotter stars since it occurs at very high temperatures $\rm > 30\,000\,K$ \citep{zirin88,lamzin89}. This implies that a continuum source at $\sim 20\,000$\, K cannot produce the \ion{He}{II}\,4686\,\AA\ line. The short duration of the line appearance indicates that such high temperatures are only present during a relatively short time.
As discussed by \citet{zirin88}, strong He II lines are observed in very hot prominences. In such prominences, the ratio of $\rm Ly \alpha$ to $\rm H{\alpha}$ is close to unity. This makes it possible to relate the $\rm H{\alpha}$ flux to the UV radiation. We also detected several other chromospheric lines as discussed in section \ref{chrom}.
\subsubsection{Identification of other chromospheric lines}
\label{chrom}
From the spectra of F2 and F3 in which we observed the \ion{He}{II}\,4686\,\AA\ line, we produced a list of the emission lines observed during the flares in addition to the three lines under analysis. Flares F2 and F3 are hotter than flare F1. For this reason, F1 does not contain any additional lines that were not already detected in F2 and F3. For example, the \ion{He}{II}\,4686\,\AA\ line is only seen in F2 and F3 not in F1. The list of the lines is presented in Table\,\ref{tab3}. Some of these lines have been observed during flares on other stars \citep[e.g.][]{montes99, montes04, paulson06,fuhr11}. In Table\,\ref{tab3}, we also present the non-flare parameters of the lines that are also observed in quiescence using the pre-flare spectrum of the night of 2018-09-19.
 The \ion{Mg}{I} lines are also present only during the flares. Several iron and helium lines are detected in these spectra, most of them occuring only at the flare peak. The observed increase in  the \ion{Fe}{I} lines suggests that the photosphere was heated during the flares as has been observed on the Sun \citep{jan18}.
\begin{table}
\caption{\label{tab3} Other chromospheric emission lines detected during flares on EV\,Lac. The wavelengths reported are the observed wavelengths in air as adopted from \citet{moore45} and \citet{nist}. N represents the energy of the lines in quiescence emitted in 600\,s.}
  \begin{tabular}{cccccc}
\hline
\hline
Ion&Wavelength ({\AA})&Multiplet&\multicolumn{3}{c}{Energy ($10^{28}$ erg)} \\
&&&F2&F3&N\\[0.5ex]
\hline
\ion{Fe}{I} 
&5269.54&15&13.06&-&-\\
&5270.36&37&7.72&-&-\\
&5318.05&406&4.89&4.84&4.05\\
&5328.04&15&7.95&-&-\\
&5371.49&15&7.72&-&-\\
&5793.93&1086&7.93&5.31\\
\ion{Fe}{II} &4549.467&38&10.03&5.70&-\\
&4583.829&38&16.28&8.68&-\\
&4629.34&37&4.98&-&-\\
&4666.75&37&3.51&-&-\\
&4923.921&42&36.87&31.0&-\\
&5018.434&42&24.24&19.85&-\\
&5169.03&42&32.55&16.35&5.08\\
&5197.57&49&7.60&-&-\\
&5234.62&49&11.51&-&-\\
&5275.97&49&9.58&-&-\\
&5316.609&49&22.01&16.44&-\\
&5325.39&49&3.11&-&4.79\\
&5362.86&48&8.80&7.42&6.93\\
&6238.38&74&1.96&1.56&-\\
&6247.56&74&2.34&2.57&-\\
&6506.33&blend&37.93&39.53&37.12\\
\ion{He}{I}&4921.9313&48&12.32&13.88&6.68\\
&4713.14, 4713.34&12&7.92&9.66&-\\
&5015.678&4&17.05&17.37&9.86\\
&5875.65&11&41.30&28.61&23.81\\
&6678.151&46\\
&7065.04&10\\
\ion{He}{II}&4685.8308&1&23.14&15.87&-\\
\ion{Mg}{I}&5167.321&2&19.32&7.75&-\\
&5172.684&2&15.83&8.03&-\\
&5183.604&2&16.55&6.57&-\\
\ion{Na}{I}&5889.85&1&16.51&7.34&4.65\\
&5895.92&1&16.04&6.41&3.55\\
\ion{V}{II}&4928.62&29&9.97&15.69&12.28\\
\ion{Ti}{II}&4571.971&82&5.47&-&-\\
\ion{H}{I} &4861.332&1&243.60&226.35&187.41\\
&6562.817&1&1324.79&1044.04&868.10\\
\hline
%\hline
\end{tabular}
~\\

\end{table}

It is possible to derive several properties of the flares from these lines. Detection of the \ion{He}{II} requires temperatures above 30\,000 K and  has so far been detected only in hot stars \cite[e.g.][]{groh08}. 
Different  diagnostics can be used to distinguish flares from prominences using the ratios of the lines. \citet{waldmeier51} classified prominences according to the strength of the magnesium lines to the \ion{Fe}{II}\,5169 line. However, according to this classification, active prominences and flares would fall in the same class and are thus not easily distinguished.

According to \citet{tand63}, limb flares can be distinguished from active prominences using the ratio of the  \ion{Fe}{II}\,5169\,\AA\, line and \ion{Ti}{II}\,4572. In flares, the intensity in the \ion{Ti}{II}\,4572 $\le $ \ion{Fe}{II}\,5169 and the \ion{Fe}{II} lines of the multiplet no. 42 gets stronger than in the other lines in the multiplets of 37 and 38. This is clearly observed from the line fluxes presented in Table\,\ref{tab3}. 
\subsection{Photometric observations}
\subsubsection{Light curve and flare identification}
\label{lightcurves}
\begin{figure}
% \subfigure
% {\includegraphics[scale=0.5]{/home/priscy/Desktop/PHD/Data/EVLac/Evlac_tesslc}}
%  \vskip -0.30cm
%  \subfigure
 {\includegraphics[scale=0.55]{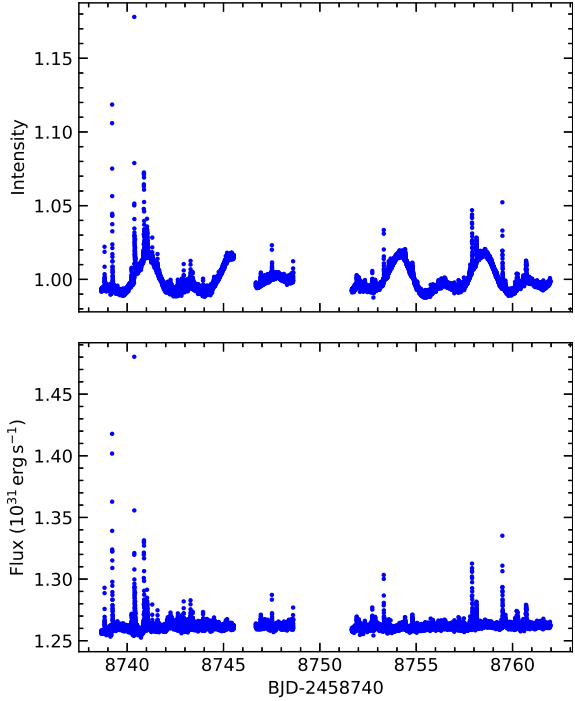}}
\caption{\label{fig4}Light curves of EV Lac. The top panel shows the modulation of the light due to starspots while the lower panel is the flux calibrated lightcurve.}
\label{lines}

\end{figure}

The TESS lightcurve of EV\,Lac shows a modulation that we interpret to be due to two starspots (see Fig.\,\ref{fig4}) located at opposite longitudes as suggested by \citet{morin08}. It is possible to estimate the properties of these spots following the method of \citet{notsu19}.
The temperature of the spots, $T_{\rm spot}$ can be determined from
\begin{equation}
 T_{\rm spot}=T_{\rm star} - 3.58\times10^{-5}T^2_{\rm star}-0.249T_{\rm star}+808,
 \label{spot}
\end{equation}
where $T_{\rm star}$ is the temperature of the star.
Using the temperature of the star from Gaia (3\,742\,K), and Eq.\,\ref{spot}, we obtained a spot temperature of $\rm \sim3\,120\,K$. It is then possible to estimate n, the ratio of the temperature of the spot to that of the star. We obtained a value of n = 0.833 which is in agreement with the range of values considered by \citet{jackson13}. The spot area can then be calculated using 
\begin{equation}
 A_{\rm spot} = \frac{\Delta F}{F}A_{\rm star}[1-n^4]^{-1},
\end{equation}
where $\frac{\Delta F}{F}$ is the normalised amplitude variation of the brightness due to the spot. We obtain a spot area of $\approx 4.8\%$ of the area of the star. This agrees well with previous studies on spot coverage on other active M dwarfs \citep[e.g.][]{Howard2019}. 

Using Fourier fitting with \textit{Period04} \citep{lenz05}, we obtained a frequency of $0.233\pm0.001\,\rm d^{-1}$  which translates into a rotation period of $4.297\pm0.017$ days, in good agreement with previous studies 4.375 days \citep{pettersen84} and 4.378 days  \citep{pettersen80, roizman84}. 

A least squares model, F(t) of the form
\begin{equation}
 F(t) = z+A_i\sum_{i}2\pi\omega_i+\phi_i,
\end{equation}
where $z$ is the zero point, $A$ the amplitude, $\omega$ the frequency and $\phi$ the phase all obtained using \textit{Period04} was fit to the data and used to normalise the lightcurve (see Fig.\,\ref{fig4}).

\begin{figure}
 \includegraphics[width=.48\textwidth]{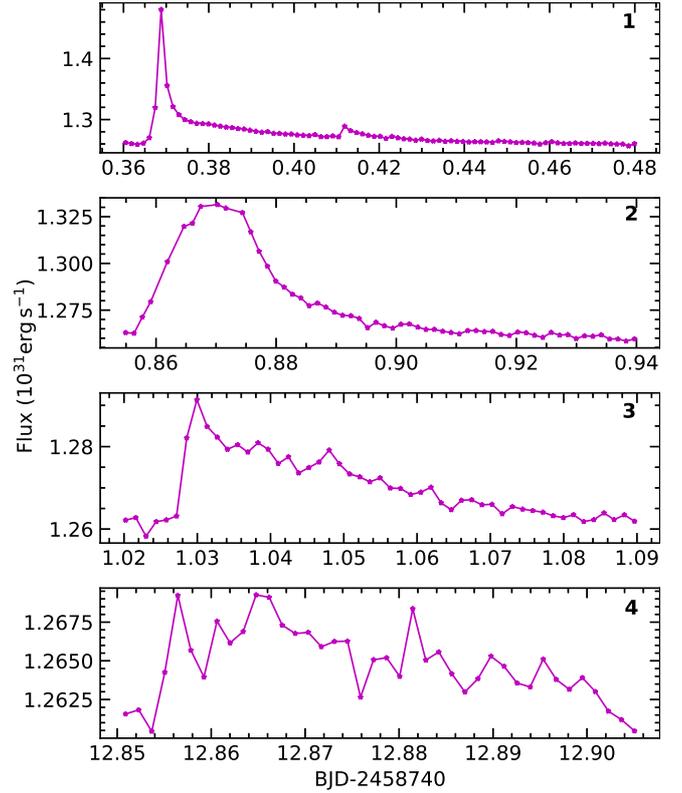}
\caption{\label{fig5} Some of the flare lightcurves that show the diverse nature of flare lightcurves on EV\,Lac. Panel 1 is the lightcurve of the largest flare that was detected from the TESS observations.}
\end{figure}

Flares in the lightcurve were identified by visual inspection. 49 flares were identified with some showing fast increase and gradual decay whereas others show a slow rise and decay features as seen in Fig.\,\ref{fig5}. 
These two types of flares usually referred to as fast (e.g. panel\,1, Fig. \ref{fig5}) and slow (e.g. panel\,2, Fig. \ref{fig5})  flares, respectively have previously been observed on this and other stars \citep[e.g.][]{dal10,dal12,kowalski11,davenport14}. White-light fast flares have been described as flares which have rise times less than 10 minutes and slow flares as those whose rise time can last $\sim$ 30 minutes \citep{dal10}. Fast flares are usually characterised with higher energy as compared to  slow flares. According to \citet{gurzadian88}, slow flares are dominated by thermal emission processes whereas in their counterparts, the dominant source of energy are non-thermal processes. \citet{dal10} found that for EV\,Lac, $75\%$ of the white-light flares are slow events while $25\%$ are fast flares. Using the classification of \citet{dal10}, $\sim 47\%$ of the flares we detected can be classified as slow while $\sim 16\%$ were fast events. The rest ($37\%$) could not easily be classified and so we term them complex flares as suggested by \citet{davenport14}.

We evaluated the connection of starspots to occurence of flares by phase folding the lightcurve using the obtained rotation period ($4.297\pm0.017$ days). We used the following ephemeris to fold the lightcurve:
\begin{equation}
 BJD = 2458738.65769+4.2969E.
\end{equation}
\begin{figure}
 \includegraphics[width=.45\textwidth]{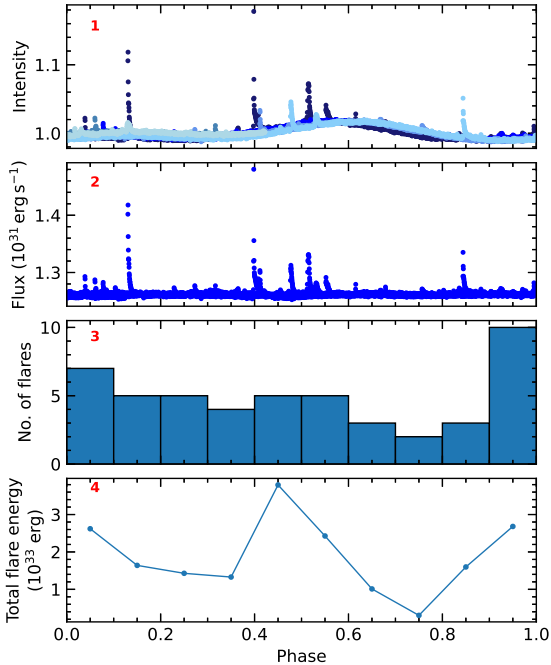}
\caption{\label{fig6} Panel 1 shows the phase folded lightcurve. The different colours represent different cycles and the colours brighten with BJD. Panel 2 is the flux calibrated phase folded lightcurve. Panel 3 is a histogram showing the flare occurence of flares in different phases. Panel 4 shows the variation of the total flare energy per bin.}
\end{figure}
Flares were seen to occur at all phases as can be seen in Fig.\,\ref{fig6}. This can be related to the presence of polar spots on this star which may influence the occurrence of flares. According to \citep{doyle18}, these spots may interact with multiple spotgroups which lead to constant flaring on this and other M stars at all rotational phases. However, it was observed that the frequency of flares  with low energy is higher at the lightcurve minimum at phase 0.9 where surface spottedness is high. This is in agreement to findings by \citep{roettenbacher18}. They suggest that weak flares appear to occur more often around the starspots probably because they are not strong enough to be seen over the stellar limb. On the other hand, we observed some strong flares at lightcurve minimum but most appear towards the lightcurve maximum (phases 0.4-0.6). This is similar to the patterns reported by \citep{vida16,vida17} on V374\,Peg and Trappist-1 objects which are both fully convective M stars. Therefore there is no one-to-one correspondence between brightness of the star and the flare activity, the situation is more complicated. This means that flares do not necessarily come from the region which contains the largest spots, because also the topology of the spot region plays a role for the flare activity.
%, but more frequently between the 0.4 to 0.6 phase which is close to the phase folded lightcurve maximum as can be seen in Fig.\,\ref{fig6}.  

\subsubsection{Flare energies}
The flare energies were determined by integrating the luminosity over the whole flare duration. To obtain the luminosity, we multiplied the normalised flux by the quiescent stellar luminosity. The flux of EV\,Lac in the TESS band was determined from the brightness of the star in the photometric bands convolved with the TESS response function.
% The quiescent luminosity of EV\,Lac was estimated using the spectrum of Vega which was scaled to the brightness of EV\,Lac. The scaled spectrum was then convolved with the TESS response function and integrated over the whole wavelength of the TESS bandpass to 
We obtained the intensity, $\rm I_{\star}= 4.135\times 10^{-9}\,erg\,s^{-1}\,cm^{-2}$. The observed luminosity was then estimated using the distance to EV\,Lac as $\rm 1.262\times10^{31}\, erg\,s^{-1}$. 

\begin{figure}
 \includegraphics[width=.52\textwidth]{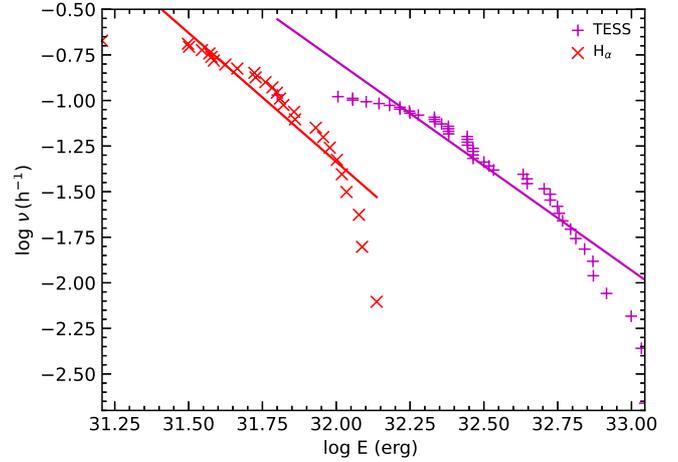}
\caption{\label{fig7} Cumulative flare frequency distribution of flares observed by TESS in comparison with those observed in $\rm H{\alpha}$.}
\end{figure}

As discussed in section \ref{flarene}, flares follow a power law distribution. In Fig.\,\ref{fig7} we show the cumulative flare distribution of flares from the TESS observations in comparison with those observed in $\rm H{\alpha}$. We obtain $\rm \beta = -1.78\pm0.17$ and $-2.63\pm0.82$ from the linear fit and the maximum likelihood estimation respectively. Previous photometric studies of EV\,Lac give values quite lower than those obtained here e.g, \citep[$-0.60\pm0.11$,~][]{lacy76},  \citep[$-0.69\pm0.12$,~][]{leto97}. However, our value of $\beta$ is in good agreement with the values ($\beta \sim -2.05$) obtained by \citet{ilin19} for stars of spectral types M3.5 - M5.5 in open clusters. Similarly, \citet{howard19} obtained a value of $\beta =-1.25$ for active M3 dwarfs. All these values are obtained from a linear fit to the flare frequency distribution. The divergence in the values from the former studies  could be due to a better sampling of high energy events in this work as compared to the previous studies since the passband used may influence the detection threshold.  
We also calculated the ratio between the continuum flux in the TESS band and the flux in the  $\rm H{\alpha}$ line to be $10.408\pm0.026$. This is in agreement with previous studies \citep[e.g.][]{somov92} who estimate a factor of 10 between the optical continuum and the $\rm H{\alpha}$ line fluxes for solar flares. We acknowledge that TESS is red sensitive and since the continuum source is $\sim 10\,000 - 20\,000$\,K, only $20-30\%$ of the total continuum emission is in the TESS band.
% We compared with flares in $\rm H{\alpha}$ because the TESS band covers the wavelength region in which $\rm H{\alpha}$ lies. This implies that photometric flares are $\sim 11$ times as energetic as the those in $H_{\alpha}$. 
\section{discussion}
%\subsection{Observations of CMEs and their constraints}
As discussed in section\,\ref{ep}, we observed several asymmetries in the Balmer lines. These can possibly be interpreted as due to flare plasma motions or erupting filaments at the impulsive phase of the flare. If we consider the latter explanation, we note that we did not observe the fast events that would be associated with them. Instead the velocities reduce as the flare evolves.
%we did not observe the fast events associated with them.

Previous studies \citep[e.g.][]{alvarado18, alvarado19} have suggested that the fast events could be suppressed by the strong magnetic fields on these active stars and thus material does not leave the surface of the star. These kind of eruptions are known as failed eruptions since the magnetic field lines do not open up and the filament bounces back.

On another hand, \citet{odert20} aimed to establish the potentially observable CME rates for stars of different spectral types and the minimum required observing time to detect stellar CMEs in Balmer lines. They suggested that CME detection is favoured for mid- to late-type M dwarfs because they have a larger fraction of the observable to intrinsic CMEs. However, these stars may require longer observing times >100 hours to detect at least one CME. We observed EV\,Lac for 127 hours which could have increased our probability of detecting atleast one CME, however these were not continuous observations and thus presenting a constraint.

CMEs expand during propagation and this leads to a decrease in the density of the plasma which could limit the duration during which a signal would be detected. This implies that if the density drops by a large factor, the CMEs become optically thin very quickly to be detected. This could also explain why we do not observe the fast events even if they managed to escape the large scale magnetic field.

We have studied the properties of flares and CMEs on EV\,Lac which is viewed equator-on and on AD\,Leo in \citetalias{priscilla20} which is viewed nearly pole-on since both stars are similar in activity. 
We note that the viewing angle of the star might not be a constraint in observing CMEs on these stars since the properties of the observed asymmetries are similar. 
% \subsection{Photoionisation model for the flares on EV\,Lac}
% We consider flares as a source of radiation which ionises a cloud of gas in a loop around it. with a luminosity of $\sim 10^{32}$ at a distance, r $\sim 10^10 cm$ from a loop of material that is ionised. In order to produce the observed spectra during flares, we performed a photoionisation model using CLOUDY \citep{ferland17} and assume the flare as an x-ray source with luminosity of $\sim 10^32\,\rm erg\,s^{-1}$ at $\sim 20$\,MK 
% %which corresponds to the temperature of the plasma in the corona \citep{robrade15}. 
% We varied the densities of the plasma in the ranges $10^8 - 10^14 \rm cm^{-3}$ inorder to reproduce the observed fluxes in $\rm H\alpha$, $\rm H\beta$ and the He\,I\,5876\,\AA line.  Several studies\citep[e.g.][]{} have modelled flares in active M dwarfs to obtain the physical parameters including electron densities and temperature, however these focus more on the Balmer lines except \citet{hawley92} who models other including Drake80, Hawley92,jevermovic98 have developed models for flares on active M dwarfs to determine their physical properties. 
\section{conclusions}
 We have studied the properties of flares and CMEs on EV\,Lac and on AD\,Leo in \citetalias{priscilla20} because both stars are similar in activity however with different viewing angles. When we observed EV\,Lac, it was a factor of 3 less active than AD\,Leo. From the spectroscopic and photometric studies of EV Lac, we  detected  27 flares (0.21 flares per hour) in $\rm H{\alpha}$ in the energy range $1.61\times 10^{31}\,-\,1.37\times 10^{32}$ erg and 49 flares ($\approx$ 0.11 flares per hour) from the TESS lightcurve with energies of $6.32\times 10^{31}\,-\,1.11\times 10^{33}$ erg.
A rotation period of $4.297\pm0.017$ days was obtained from the TESS lightcurve. Most of the flares show a complex lightcurve both in  spectroscopic and photometric observations and no direct correlation between flaring and  starspots was found. Statistical analysis between the energies of the flares observed with TESS and those obtained from the spectroscopic observations shows that the ratio of the continuum flux in the TESS passband to the energies emitted in  $\rm H \alpha$ is $10.408\pm0.026$.

Power-law index $\beta$ values in the range $-1.78\pm0.17\,-\, -2.58\pm0.51$ from a linear fit of the cumulative flare frequency distribution and $-1.96\pm 0.55\,-\, -2.71\pm0.94$ using the maximum likelihood estimation were obtained.
Three white-light flares were detected in the spectra of EV\,Lac. From these flares, it was determined that the area of the emitting region is very small ($\le 0.07\%$) of the area of the star. At the peak of two of these flares, we also detected the \ion{He}{II}\,4686 \AA\, line which is known to occur only in very hot plasma. Its short appearance at the flare peak indicates that the high plasma temperatures are present only for a short time. Additional chromospheric lines present during the flares were also identified.

We also analysed the spectra by studying the behaviour of the chromospheric lines $\rm H{\alpha}$, $\rm H{\beta}$, and \ion{He}{I}\,$\rm D_3$  to detect CMEs. We observed asymmetries in several spectra nearly all of them showing both blue and red wing asymmetries. All the blue wing asymmetries have velocities much smaller than the escape velocity of the star. These slow velocity events may be indicative of the rarity of CMEs on these stars. However, in one relatively weak event, we found an asymmetry in $\rm H\alpha$ of $\sim 220\,\rm km\,s^{-1}$ which we interpret as a signature of a quiet region erupting filament. 
We however note that the fast event associated with the erupting filament was not detected even if we assume that it could have originated from a quiet region. 
From these asymmetries, we conclude that the viewing angle may not constrain the detection of CMEs on M dwarfs since the properties of the  asymmetries observed in EV\,Lac are similar to those on AD\,Leo.

The slow events could be due to suppression by the magnetic field of the star and hence they are not accelerated to high speeds. On the other hand, it could also be that we do not detect the fast events because as the CME expands during propagation, the signal is diluted.
It still remains unclear if actually CMEs are rare on this and other active M dwarfs or it is just that we are constrained by the observation strategies. 
\section*{Acknowledgements}
The authors are grateful for the funding from the International Science Program at Uppsala University. The authors greatly acknowledge the  Th\"{u}ringer Landessternwarte Observatory, Germany for giving them observation time. We are also grateful to Robert Greimel, Petra Odert and Martin Leitzinger for the helpful discussion. We thank the referee for their helpful comments.

This work was generously supported by the Th\"{u}ringer Ministerium f\"{u}r Wirtschaft, Wissenschaft und Digitale Geselischaft. 
This manuscript includes data collected with the TESS mission obtained from the MAST data archive at the Space
Telescope Science Institute (STScI). Funding for the TESS mission is provided by the NASA Explorer Program. STScI is operated by the Association of Universities for Research in Astronomy, Inc., under NASA contract NAS 526555.
This research made use of the SIMBAD database, operated at CDS, Strasbourg, France. This research also made use of data from the European Space Agency (ESA) mission \textit{Gaia} (\url{https://www.cosmos.esa.int/gaia}), processed by the \textit{Gaia} Data Processing and Analysis Consortium (DPAC) (\url{https://www.cosmos.esa.int/web/gaia/dpac/consortium}).

\section{Data availability}
The spectroscopic data underlying this article will be shared on reasonable request to the corresponding author.
The TESS data are publicly available at the Mikulski Archive for Space Telescopes (MAST) site via \url{https://archive.stsci.edu/tess/}.
%%%%%%%%%%%%%%%%%%%%%%%%%%%%%%%%%%%%%%%%%%%%%%%%%%



\bibliographystyle{mnras}
\bibliography{paper} % if your bibtex file is called example.bib

% \begin{thebibliography}{99}
% \bibitem[\protect\citeauthoryear{Author}{2012}]{Author2012}
% Author A.~N., 2013, Journal of Improbable Astronomy, 1, 1
% \bibitem[\protect\citeauthoryear{Others}{2013}]{Others2013}
% Others S., 2012, Journal of Interesting Stuff, 17, 198
% \end{thebibliography}

%%%%%%%%%%%%%%%%%%%%%%%%%%%%%%%%%%%%%%%%%%%%%%%%%%

%%%%%%%%%%%%%%%% APPENDICES %%%%%%%%%%%%%%%%%%%%%

\appendix

\section{observation log}

\begin{table*}

\caption{\label{tab1}Observation log showing nights of observation of EV\,Lac; the number of hours per night and the corresponding number of spectra obtained are shown in columns 3 and 4 respectively. }
  \begin{tabular} {p{1.7cm} p{3.0cm} p{1.5cm} p{1.5cm} p{1.5cm} p{0.05cm}cccccc}
\hline
\hline
Obs. date & Start--End time (UTC) & Obs  (h) & Number of Spectra \\

%[0&.5ex]\cline{6-7}
% (H$_\alpha$)	(H$_\beta$) [0.5ex]
\hline
2016-10-10 &23:10 - 23:30 &0.33&2\\
2016-10-16 &17:50 - 00:31 &6.33&38\\
2016-11-14 &17:58 - 22:32 &2.0 &12\\
2016-11-18 &20:21 - 23:51 &3.33&20\\
2016-11-19 &18:10 - 23:57 &3.83&23\\
2016-11-20 &17:19 - 23:26 &5.83&35\\
2016-12-16 &18:14 - 22:23 &2.67&16\\
2017-12-26 &18:34 - 19:26 &0.83&5\\
2018-01-04 &21:14 - 21:25 &0.17&1\\
2018-01-05 &20:32 - 20:51 &0.33&2\\
2018-08-21 &22:00 - 02:36&4.33&26\\
2018-08-22 &19:17 - 02:35 &6.33&38\\
2018-08-28 &20:42 - 03:12 &6.17&37\\
2018-09-17 &21:07 - 03:24 &6.0&36\\
2018-09-18 &21:30 - 02:44 &5.0&30\\
2018-09-19 &20:52 - 03:23 &6.0&36\\
2018-10-16 &19:55 - 00:21 &3.0&18\\
2018-10-17 &17:28 - 00:59 &7.17&43\\
2018-10-20 &17:27 - 21:00 &2.83&17\\
2018-10-21 &18:30 - 00:48 &6.0&36\\
2019-08-01 &22:25 - 23:29 &1.33&8\\
2019-08-02 &21:17 - 21:27 &0.17&1\\
2019-08-03 &00:24 - 01:36 &0.83&5\\
2019-08-04 &21:52 - 22:35&0.67&4\\
2019-08-05 &22:40 - 01:00 &2.17&13\\
2019-08-07 &21:20 - 02:13 &4.67&28\\
2019-08-08 &20:58 - 01:30 &3.83&23\\
2019-08-10 &23:23 - 02:09 &2.50&15\\
2019-08-12 &00:41 - 02:23 &1.50&9\\
2019-08-13 &20:09 - 23:24 &3.0&18\\
2019-08-29 &00:27 - 02:44 &2.33&14\\
2019-08-30 &19:42 - 02:32&6.50&39\\
2019-08-31 &21:29 - 01:41&4.0&24\\
2019-09-02 &20:45 - 03:03&6.0&36\\
2019-09-03 &20:42 - 02:26&5.33&32\\
2019-09-04 &20:34 - 00:04&3.33&20\\
\hline
%\hline
\end{tabular}

\end{table*}

%%%%%%%%%%%%%%%%%%%%%%%%%%%%%%%%%%%%%%%%%%%%%%%%%%

% Don't change these lines
\bsp	% typesetting comment
\label{lastpage}
\end{document}